\documentclass[fleqn,usenatbib]{mnras}

\usepackage{newtxtext,newtxmath}

\usepackage[T1]{fontenc}
\usepackage[usenames,dvipsnames,svgnames,table]{xcolor}
\usepackage{comment}
\usepackage{subcaption}
\usepackage{multirow}
\usepackage[flushleft]{threeparttable}

\DeclareRobustCommand{\VAN}[3]{#2}
\let\VANthebibliography\thebibliography
\def\thebibliography{\DeclareRobustCommand{\VAN}[3]{##3}\VANthebibliography}

\newsavebox{\measurebox}

\usepackage{graphicx}	
\usepackage{amsmath}	
\usepackage{scalerel,tikz}
\usetikzlibrary{svg.path}
\definecolor{orcidlogocol}{HTML}{A6CE39}
\tikzset{orcidlogo/.pic={
		\fill[orcidlogocol] svg{M256,128c0,70.7-57.3,128-128,128C57.3,256,0,198.7,0,128C0,57.3,57.3,0,128,0C198.7,0,256,57.3,256,128z};
		\fill[white] svg{M86.3,186.2H70.9V79.1h15.4v48.4V186.2z}
		svg{M108.9,79.1h41.6c39.6,0,57,28.3,57,53.6c0,27.5-21.5,53.6-56.8,53.6h-41.8V79.1z M124.3,172.4h24.5c34.9,0,42.9-26.5,42.9-39.7c0-21.5-13.7-39.7-43.7-39.7h-23.7V172.4z}
		svg{M88.7,56.8c0,5.5-4.5,10.1-10.1,10.1c-5.6,0-10.1-4.6-10.1-10.1c0-5.6,4.5-10.1,10.1-10.1C84.2,46.7,88.7,51.3,88.7,56.8z};
}}
\newcommand\orcidicon[1]{\href{https://orcid.org/#1}{\mbox{\scalerel*{
				\begin{tikzpicture}[yscale=-1,transform shape]
					\pic{orcidlogo};
				\end{tikzpicture}
			}{|}}}}

\usepackage{hyperref}

\newcommand{\meraxes}{\ifmmode\mathrm{\textit{Meraxes}}\else{}{\textit{Meraxes} }\fi}

\newcommand{\Msun}{\ifmmode\mathrm{M_\odot}\else{}$\rm M_\odot$\fi}

\newcommand{\xH}{\ifmmode{x}_{\rm HI}\else{}${x}_{\rm HI}$\fi}

\title[JWST $z{\gtrsim}12$ analogues]{Implications of $z{\gtrsim}12$ JWST galaxies for galaxy formation at high redshift}

\author[Qin et al.]{Yuxiang Qin~\orcidicon{0000-0002-4314-1810}$^{1,2}$\thanks{E-mail: Yuxiang.L.Qin@gmail.com}, Sreedhar Balu~\orcidicon{0000-0002-5281-5151}$^{1,2}$ 
	and J. Stuart B. Wyithe~\orcidicon{0000-0001-7956-9758}$^{1,2}$
\\
$^{1}$School of Physics, University of Melbourne, Parkville, VIC 3010, Australia\\
$^{2}$ARC Centre of Excellence for All Sky Astrophysics in 3 Dimensions (ASTRO 3D)\\
}

\date{\today}

\pubyear{2022}

\begin{document}
\label{firstpage}
\pagerange{\pageref{firstpage}--\pageref{lastpage}}
\maketitle

\begin{abstract}
	Using a semi-analytic galaxy-formation model, we study analogues of 8 recently discovered JWST galaxies at $z{\gtrsim}12$. We select analogues from a cosmological simulation with a $(311{\rm cMpc})^3$ volume and an effective particle number of $10^{12}$ enabling resolution of every atomic-cooling galaxy at $z{\le}20$. We vary model parameters to reproduce the observed UV luminosity function at $5{<}z{<}13$, aiming for a statistically representative high-redshift galaxy mock catalogue. Using the forward-modelled JWST photometry, we identify analogues from this catalogue and study their properties as well as possible evolutionary paths and local environment. 	
	We find faint JWST galaxies ($M_{\rm UV}{\gtrsim}-19.5$) to remain consistent with standard galaxy-formation model and that our fiducial catalogue includes large samples of their analogues. The properties of these analogues broadly agree with conventional SED fitting results, except for having systematically lower redshifts due to the evolving UV luminosity function, and for having higher specific star formation rates as a result of burstier histories in our model. On the other hand, only a handful of bright galaxy analogues can be identified for the observed $z{\sim}12$ galaxies. Moreover, in order to reproduce the $z{\gtrsim}16$ JWST galaxy candidates, boosted star-forming efficiencies and reduced feedback regulation are necessary  relative to models of lower-redshift populations. This suggests star formation in the first galaxies could differ significantly from their lower-redshift counterparts. We also find that these candidates are subject to low-redshift contamination, which is present in our fiducial results as both the dusty or quiescent galaxies at $z{\sim}5$.
\end{abstract}




\section{Introduction}
If the Hubble Space Telescope (HST) gave us a glimpse of the $z{\gtrsim}10$ Universe through the keyhole, JWST has undoubtedly opened the door. Since its Early Release Observations (ERO; \citealt{Pontoppidan2022ApJ...936L..14P}) and Director’s Discretionary Early Release Science (ERS) programs were made publicly available, a large number of $z{\gtrsim}10$ candidates have been reported by various independent groups using NIRCam imagings \citep{Naidu2022ApJ...940L..14N,Naidu2022arXiv220802794N,Castellano2022ApJ...938L..15C,Yan2023ApJ...942L...9Y,Finkelstein2022ApJ...940L..55F,Atek2023MNRAS.519.1201A,Donnan2023MNRAS.518.6011D,Whitler2020MNRAS.495.3602W,Harikane2023ApJS..265....5H,Zavala2023ApJ...943L...9Z,Rodighiero2022arXiv220802825R,Adams2022arXiv220711217A,Cullen2022arXiv220804914C,Furtak2023MNRAS.519.3064F,Ono2022arXiv220813582O,Bradley2022arXiv221001777B,Robertson2022arXiv221204480R}. These preliminary results 
include some discoveries which challenge the current concordance model \citep{BK2009MNRAS.398.1150B,Haslbauer2022ApJ...939L..31H,Parashari2023arXiv230500999P}. Following up on the recent NIRSpec result \citep{Curtis-Lake2022arXiv221204568C} pushing the earliest, {\it spectroscopically confirmed} high-redshift galaxy to $z=13.2_{-0.07}^{+0.04}$, we aim to quantify whether standard galaxy formation models are consistent with these observables at the cosmic dawn.

According to the standard galaxy-formation model (see reviews by \citealt{Somerville2015,Naab2017ARA&A..55...59N,Vogelsberger2020NatRP...2...42V} and references therein), galaxies form from the gravitational collapse of over-dense regions of dark matter and gas, which subsequently grow through mergers and accretion. Gas cools and condenses to form stars, which illuminate the host galaxy and allow us to observe its complex structure. The standard model also accounts for the role of feedback processes including energy, mass and metals injected by supernovae and central supermassive black holes. These have been proven essential to regulating star formation and shaping galaxy properties.

\begin{figure*}
	\centering
	\includegraphics[width=\textwidth]{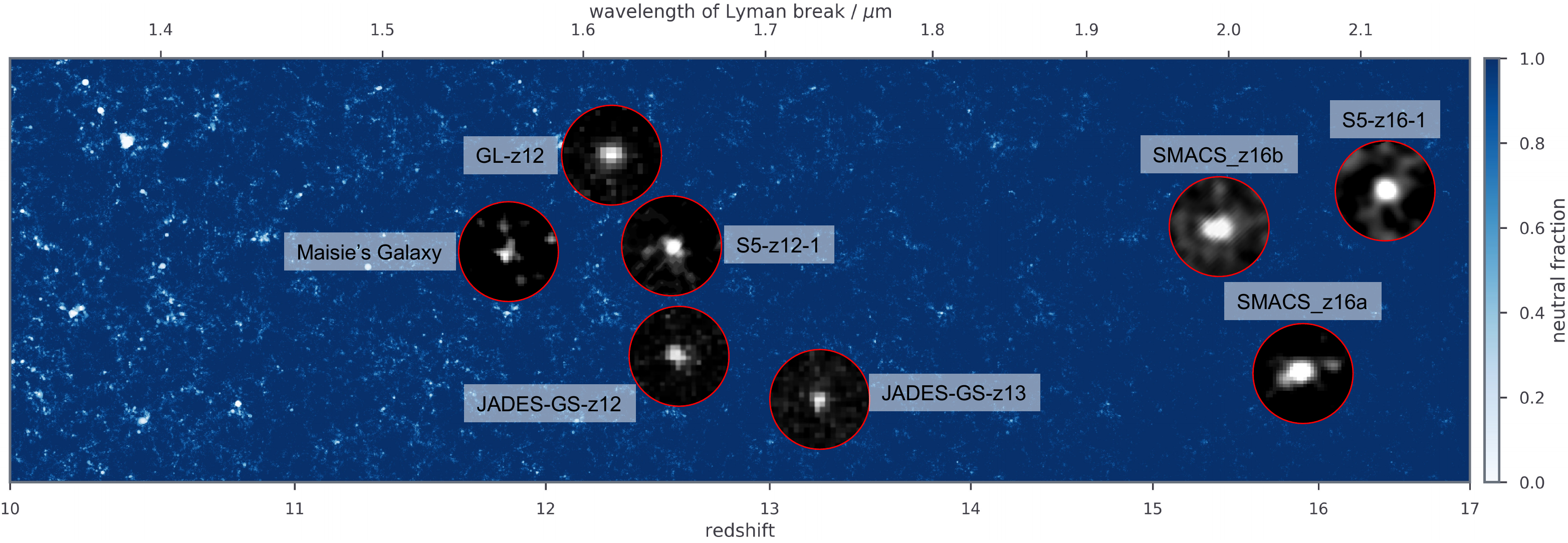}\\
	\caption{Eight $z\gtrsim12$ JWST targets explored in this work with background illustrating the early stage of reionization (projected with a depth of 4 cMpc, a typical bubble radius for high-redshift bright galaxies) according to our fiducial model. References for these galaxy observations are listed in Table \ref{tab:galaxies}.}\label{fig:lightcone}
\end{figure*}

\begin{table*}
	\centering
	\caption{Properties of the $z\gtrsim12$ JWST targets explored in this work. The last columnr states whether the galaxy is consistent with our galaxy-formation models.}
	\begin{tabular}{m{2mm}cccm{15mm}cm{29mm}m{31mm}} 
		\hline\\[-3mm]
		Sec & \ \ \ \ ID &   $z$  &   $M_{\rm UV}$ &   $\log_{10}[M_*/{\rm M}_\odot]$ &  $\rm{SFR}$[${\rm M}_\odot\ {\rm yr}^{-1}$]$^a$ &    References$^b$ &   w.r.t. our models\\
		\hline \\[-3mm]
		\ref{jadesz12} & JADES-GS-z13 & $13.20^{+0.04}_{-0.07}$ &  $-18.5\pm0.2$ & $7.8^{+0.4}_{-0.5}$ & $1.0_{-0.5}^{+1.0}$& CL23, R22, D23b, H23a  & \multirow{3}{3.1cm}{\\consistent with fiducial}  \\[1mm]
		
		\ref{jadesz12} & JADES-GS-z12 & $12.63^{+0.24}_{-0.08}$  & $-18.8\pm0.1$ & 	$8.4^{+0.4}_{-0.7}$ &$1.3_{-0.9}^{+1.9}$ & CL23, R22, H23a &   \\[1mm]
		
		\ref{S5-z12-1} & S5-z12-1 & $12.58^{+1.23}_{-0.46}$ & $-20.2\pm0.1$ & $8.53^{+0.61}_{-0.69}$ & $5.5^{+4.7}_{-4.4}$ & H23b & \\[1mm]
		\hline\\[-2mm]		
		
		\ref{gnlz12} & GLz12 & $12.2^{+0.1}_{-0.2}$  & $-21.0\pm0.1$ & $9.1^{+0.3}_{-0.4}$ & $6^{+5}_{-2}$ & N22, B23, D23a, H23b, O22, P23, S22 &  \multirow{2}{3.1cm}{consistent with fiducial but challenging} \\[1mm]
		
		
		\ref{CEERSz14} & Maisie’s Galaxy& $11.44^{+0.09}_{-0.08}$ &  $-20.32^{+0.08}_{-0.06}$ & $8.50^{+0.29}_{-0.44}$ & $2.1^{+4.8}_{-2.0}$  & FI22, AH23, D23a, H23ab, O22, Z22 &	\\[1mm]		
		
		\hline\\[-2mm]		
		\ref{subsection:smacz16ab} & SMACS\_z16a  & $15.92^{+0.17}_{-0.15}$ & $-20.59\pm0.15$  & $8.79^{+0.32}_{-0.33}$ & $16.6^{+2.9}_{-16.4}$ &   \multirow{2}{3.1cm}{AT23, AD23, FU23, H23b}  & \multirow{2}{3.1cm}{inconsistent with fiducial, requiring maxSF}	\\[1mm]
		
		\ref{subsection:smacz16ab} & SMACS\_z16b  & $15.32^{+0.16}_{-0.13}$ & $-20.96\pm0.14$  & $8.80^{+0.44}_{-0.25}$ & $57.5^{+38.0}_{-29.4}$ &  &	\\[1mm]
		\hline\\[-2mm]		
		
		
		\ref{subsection:z17} & S5-z16-1 & $16.41^{+0.66}_{-0.55}$ & $-21.6{\pm}0.3$ & $8.59^{+1.23}_{-0.31}$ & $5.1^{+21.7}_{-1.8}$ & H23b, O22 & {inconsistent with fiducial, even challenging to maxSF}\\[1mm]
		\hline\\[-2mm]		
		
		
	\end{tabular}
	\begin{tablenotes}
		\item $^a$ SFR is averaged over 50Myr except for JADES-GS-z12, JADES-GS-z13 and Maisie’s Galaxy in which 30Myr, 30Myr and 10Myr are considered, respectively.
		\item $^b$ References are \citet[][AD23]{Adams2023MNRAS.518.4755A}, \citet[][AH23]{haro2023spectroscopic}, \citet[][AT23]{Atek2023MNRAS.519.1201A},  \citet[][B23]{Bakx2023MNRAS.519.5076B}, \citet[][CL23]{Curtis-Lake2022arXiv221204568C}, \citet[][D23a]{Donnan2023MNRAS.518.6011D},
		\citet[][D23b]{Donnan2023MNRAS.520.4554D}, \citet[][FI22]{Finkelstein2022ApJ...940L..55F}, \citet[][FU23]{Furtak2023MNRAS.519.3064F},  \citet[][H23ab]{Harikane2023arXiv230406658H,Harikane2023ApJS..265....5H},
		 \citet[][N22]{Naidu2022ApJ...940L..14N}, \citet[][O22]{Ono2022arXiv220813582O}, \citet[][P23]{Popping2023A&A...669L...8P},
		 \citet[][R22]{Robertson2022arXiv221204480R},
		  \citet[][S22]{Santini2023ApJ...942L..27S},
		\citet[][Z22]{Zavala2023ApJ...943L...9Z}.
	\end{tablenotes}
	\label{tab:galaxies}
\end{table*}

Using a semi-analytic galaxy-formation model (introduced in Section \ref{sec:model}), we make realizations of the early Universe including NIRCam broad-band photometry for billions of galaxies at $z{\ge}5$. By varying efficiencies of star formation and feedback, this theoretical galaxy population is calibrated to represent summary statistics of the large sample observed before JWST, mostly at $z{\le}10$. In this modelled galaxy catalogue, we then seek those having a spectral energy distribution (SED) close to the new JWST observations, for which we include both the NIRSpec targets and several NIRCam candidates\footnote{As earlier releases of the NIRCam candidates are often revised in the final publications, we only discuss targets that are currently in press.} at $z{\gtrsim}12$. Our objective is to (1) probe the potential formation history and local environment of galaxies formed in the first ${\sim}300$Myr of our Universe using modelled analogues, where identifiable; and otherwise (2) study the implications for standard galaxy formation model where the modelled galaxies are inconsistent with observations.

The cosmic chronology of our targets is illustrated in Fig. \ref{fig:lightcone} and their inferred properties from various observational campaigns are summarized in Table \ref{tab:galaxies}. After discussing their implications for galaxy-formation models in Section \ref{subsection:glfz12}, we present analogues of these JWST galaxies in Section \ref{sec:analogues}. Section \ref{sec:conclusion} concludes our results. Cosmological parameters from Planck 2018 ($\Omega_{\mathrm{m}}, \Omega_{\mathrm{b}}, \Omega_{\mathrm{\Lambda}}, h, \sigma_8, n_\mathrm{s} $ = 0.312, 0.0490, 0.688, 0.675, 0.815, 0.968; \citealt{Planck2020A&A...641A...6P}) are adopted in this study.

\section{Modelling the first galaxies during the epoch of reionization}\label{sec:model}

\begin{figure*}
	\includegraphics[width=\textwidth]{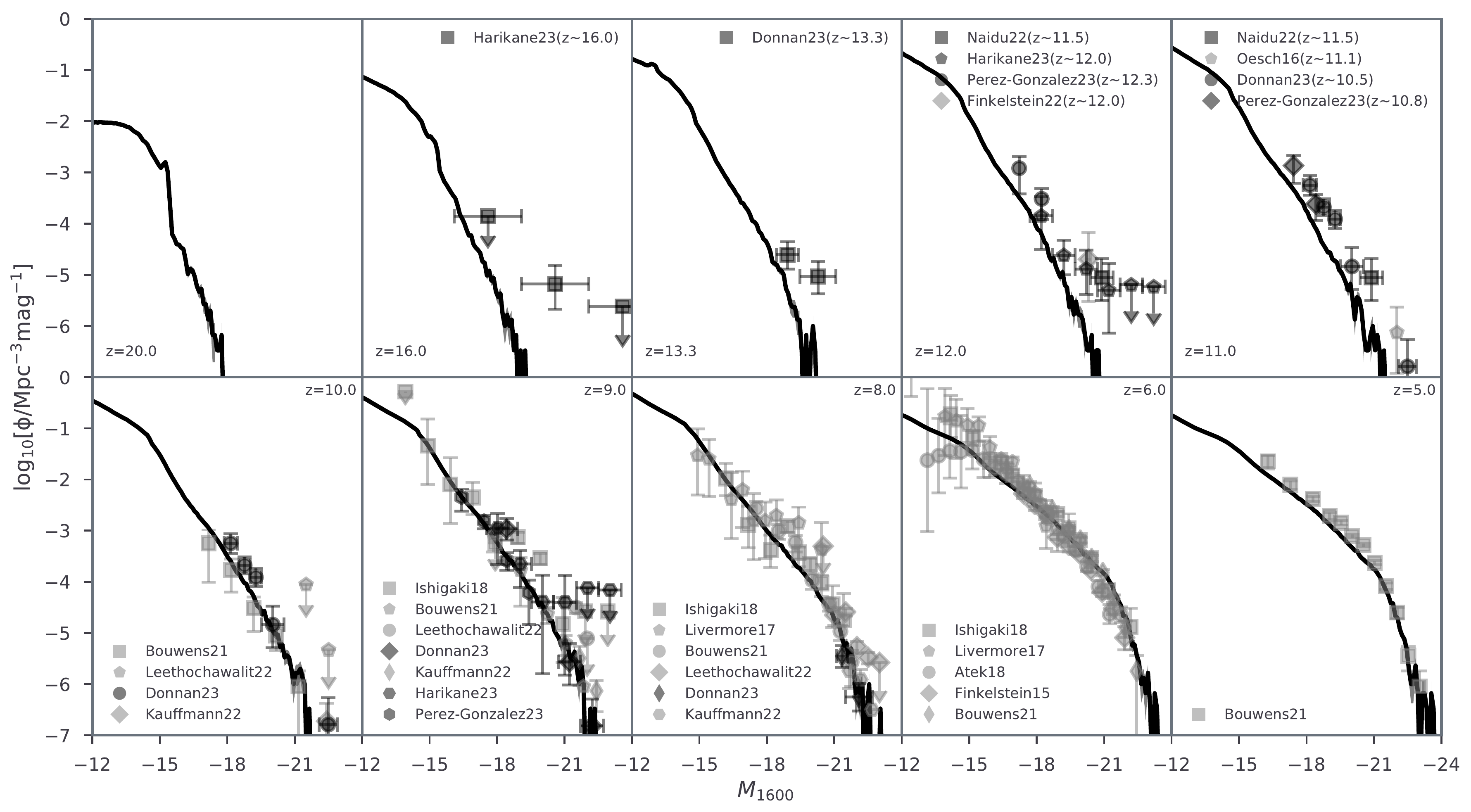}
	\caption{Galaxy UV non-ionizing luminosity functions from $z{=}20$ to 5 predicted by our fiducial model, which was calibrated to reproduce the observational data (light grey) prior to JWST such as \citet{Finkelstein2015ApJ...810...71F,Oesch2016ApJ...819..129O,Livermore2017ApJ...835..113L,Atek2018MNRAS.479.5184A,Ishigaki2018ApJ...854...73I,Bhatawdekar2019MNRAS.486.3805B,Bouwens2021AJ....162...47B,Bouwens2022arXiv221102607B,Leethochawalit2022arXiv220515388L} and \citet{Kauffmann2022arXiv220711740K}. Shaded regions and error bars indicate the $1\sigma$ Poisson error from the model and observations. We also highlight (dark grey) recent JWST results from \citet{Donnan2023MNRAS.518.6011D,Finkelstein2022ApJ...940L..55F,Harikane2023ApJS..265....5H,Naidu2022ApJ...940L..14N} and \citet{PG2023arXiv230202429P}, which are still broadly consistent with our model prediction at least in the faint range. We also note that a $z{\sim}16$ candidate (CEERS-93316) selected by \citet{Donnan2023MNRAS.518.6011D} and \citet{Harikane2023ApJS..265....5H} has recently been refuted spectroscopically. However, \citet{Atek2023MNRAS.519.1201A} reported two other $z{\sim}16$ candidates in the same field (SMACS J0723) and hence a revised number density will likely be at a similar level.}
	\label{fig:glfs}
\end{figure*}

In this work, we use the \meraxes semi-analytic model (SAM;  \citealt{Mutch2016MNRAS.463.3556M}), designed to study the Epoch of Reionization (EoR). The model is applied to dark matter halo merger trees introduced in \cite{Balu2022}, which were constructed (with the \textsc{VELOCIraptor} halo-finder and \textsc{TreeFrog} algorithm by \citealt{VELOCIRAPTOR,TREEFROG}) from an \textit{N}-body simulation of $210h^{-1}$ cMpc (performed by Power et al. in prep using SWIFT by \citealt{Schaller2023arXiv230513380S}) that has been augmented (with \textsc{DarkForest} by \citealt{DARKFOREST}) to resolve all atomic cooling halos at $z\le20$ (i.e., $M_{\rm vir}\ge3{\times}10^7{\rm M}_\odot$). With halo properties inherited from the merger trees, our SAM assigns galaxies with a baryonic component according to the cosmic mean and the strength of local photo-ionization. It then evaluates galaxy properties based on various astrophysical processes including gas accretion and cooling, stellar evolution and feedback, as well as metal enrichment, satellite infall, and merger events. 

We also incorporate standard stellar population synthesis (using the instantaneous model with nebular continuum included from \textsc{starburst99} by \citealt{Leitherer1999ApJS..123....3L}) and estimate attenuation by the interstellar (ISM) and intergalactic medium (IGM; following \citealt{Charlot2000ApJ...539..718C} and \citealt{Inoue2014MNRAS.442.1805I}) to calculate galaxy spectra as well as SEDs with the NIRCam wide-band filters. The model also includes feedback from AGN \citep{Qin2017MNRAS.472.2009Q} but their UV emission is ignored in this work. We assume 15 per cent of UV ionizing photons and X-rays above 500 eV escape from the host galaxy, and study their impact on the large-scale neutral IGM (via an excursion-set algorithm based on \textsc{21cmFAST} by \citealt{Mesinger2011MNRAS.411..955M}; see also \citealt{Murray2020JOSS....5.2582M}). Fig. \ref{fig:lightcone} illustrates the early stage of reionization according to our fiducial model, for which the late-time prediction and integrated EoR history are consistent with quasar/Lyman-alpha emitter observations (\citealt{Mcgreer2015,Wang2020,Banados2017,Greig2017,Greig2019,Greig2022,Davies2018,Wold2022ApJ...927...36W,Inoue2018PASJ...70...55I,Morales2021ApJ...919..120M,Ouchi2018PASJ...70S..13O,Whitler2020MNRAS.495.3602W,Mason2018ApJ...856....2M,Jung2020ApJ...904..144J,Qin2021MNRAS.506.2390Q,Bolan2022MNRAS.517.3263B}; Campo et al. in prep) and {\it Planck}'s latest measurement of the cosmic microwave background \citep{Planck2020A&A...641A...6P}. This fiducial model also results in a high-redshift galaxy and quasar population that is statistically representative of the observed Universe, including the predicted stellar mass function and UV non-ionizing luminosity function calibrated against observations across cosmic time ($z{\sim}5$--10; \citealt{Qin2017MNRAS.472.2009Q}). 

The adopted fiducial parameters\footnote{A further tuning was needed when the new merger trees are employed (see more in \citealt{Balu2022}).} and how we tuned them are outlined in detail by \citet[][]{Qiu19} where Bayesian inference was performed against the observed UV luminosity function and colour-magnitude relation prior to JWST. Here, we illustrate the luminosity function between $z{=}20$ and 5 in Fig. \ref{fig:glfs} with some latest measurements including those taking advantage of the early JWST data. While the model was only calibrated against observations at lower redshifts, its prediction at $z{>}10$ remains accurate with only some discrepancies at the very bright end which we further examine in the next section.

\begin{figure*}
	\includegraphics[width=0.85\textwidth]{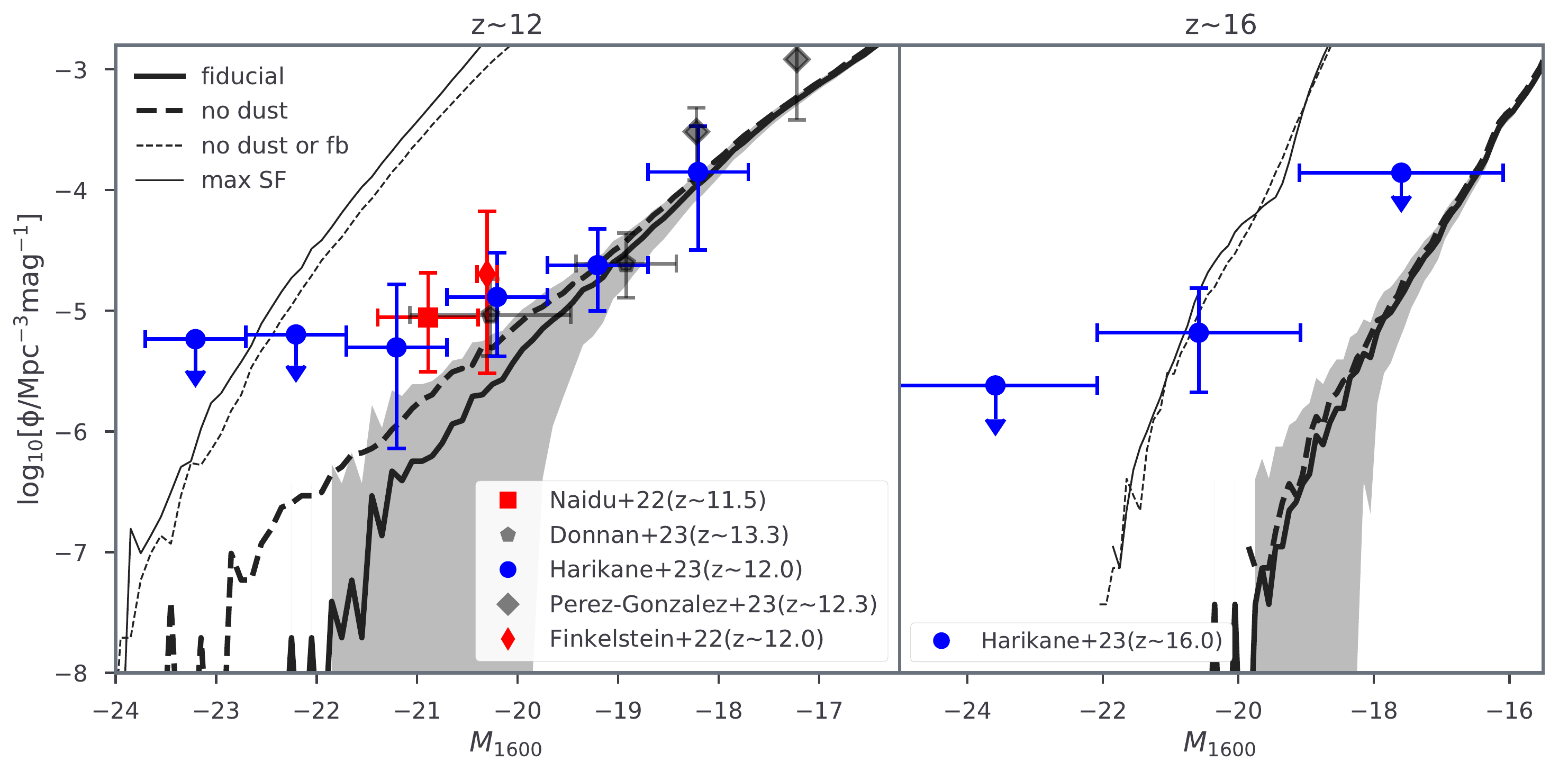}\vspace*{-1mm}
	\caption{\textit{Left panel:} similar to Fig. \ref{fig:glfs} but focused on luminosity function at redshifts around S5-z12-1 \citep{Harikane2023ApJS..265....5H}, GLz12 \citep{Naidu2022ApJ...940L..14N} and Maisie's Galaxy \citep{Finkelstein2022ApJ...940L..55F} -- 17 consecutive snapshots from {\it Meraxes} between $z{=}10$ and 13 are considered as independent realizations to reduce the sample variance for bright galaxies. In addition to the fiducial prediction, models that ignore dust attenuation ({\it no dust}), supernova feedback ({\it no dust or fb}) and maximize star formation efficiency ({\it maxSF}) are shown for comparison together with the early JWST estimates by \citet{Naidu2022ApJ...940L..14N,Donnan2023MNRAS.518.6011D,Harikane2023ApJS..265....5H,PG2023arXiv230202429P} and \citet{Finkelstein2022ApJ...940L..55F}. \textit{Right panel:} luminosity function at redshifts around $z{=}16$. Our simulation results are based on 10 consecutive snapshots between $z{=}15$ and 17 while observational estimates come from \citet{Harikane2023ApJS..265....5H}. As noted in Fig. \ref{fig:glfs}, CEERS-93316 that was considered as a $z{\sim}16$ candidate by \citet{Harikane2023ApJS..265....5H} is in fact a $z=4.9$ low star-forming dusty galaxy \citep{haro2023spectroscopic}. However, as the two $z{\sim}16$ candidates reported by \citet{Atek2023MNRAS.519.1201A} remain promising and were found in the same field as what \citet{Harikane2023ApJS..265....5H} explored, a revised number density will likely be at a similar level.
	}

\label{fig:glf_z12}
\end{figure*}

\section{Implications of bright JWST galaxies for a standard galaxy-formation model}\label{subsection:glfz12}

GLz12 \citep{Naidu2022ApJ...940L..14N} and Maisie's Galaxy \citep{Finkelstein2022ApJ...940L..55F} are two bright galaxies at $z{\sim}12$. However, observing them in these small-volume ERS programs indicates a surprisingly large number density for high-redshift bright galaxies -- GLz12 sets the number density for galaxies of $M_{\rm UV}{\sim}{-}21$ at $\phi{\sim}10^{-5}{\rm Mpc}^{-3}{\rm mag}^{-1}$ together with another $z{\sim}10$ candidate reported in GLASS \citep{Naidu2022ApJ...940L..14N}; while Maisie's Galaxy suggests $\phi{\sim}2{\times}10^{-5}{\rm Mpc}^{-3}{\rm mag}^{-1}$ at a luminosity ${\sim}1$ magnitude fainter. These values, although having large uncertainties, are consistent with each other and more recent estimates from much larger samples \citep{Donnan2023MNRAS.518.6011D,Harikane2023ApJS..265....5H,PG2023arXiv230202429P}. 

On the other hand, theoretical models that tie galaxy formation closely to their host halo growth seem to struggle to simultaneously match both the bright and faint end of the luminosity function as well as its cosmic evolution (e.g. \citealt{Dayal2014MNRAS.445.2545D,Behroozi2019MNRAS.488.3143B} used in \citealt{Naidu2022ApJ...940L..14N}; see also discussion in \citealt{Mason2022arXiv220714808M,yung2023ultrahighredshift}). This is also the case for our model. We highlight the galaxy UV luminosity function at $z{=}10$--13 from multiple snapshots of {\it Meraxes} output in the left panel of Fig. \ref{fig:glf_z12}. To facilitate discussion of possible failures of semi-analytic galaxy formation models at high redshift, we also add results from
\begin{enumerate}
	\item \textit{no dust}, a fiducial model where dust attenuation in stellar birth clouds and in the ISM is ignored to explore the possibility that extrapolating our dust model to such early Universe might be inaccurate;
	\item \textit{no dust or fb}, a \textit{no dust} model with supernova (and reionization) feedback further minimized to study scenarios where feedback in the first galaxies is much weaker than previously expected;
	\item \textit{maxSF}, a \textit{no fb} model with star formation efficiency further maximized to illustrate some of these JWST candidates might be forming stars at much higher rates compared to galaxies at $z{\lesssim}10$.
\end{enumerate}

\subsection{Model modifications to reproduce more bright galaxies at high redshift}
The dust model in our simulation is based on \cite{Charlot2000ApJ...539..718C}, linking attenuation in stellar birth clouds and the ISM to star formation rates, metallicities and gas column densities. Its parameters were chosen after a rigorous Bayesian exploration \citep{Qiu19} with constraints from UV luminosity functions and colour-magnitude relations at relatively lower redshifts ($z{\sim}4-7$; \citealt{Bouwens2014ApJ...793..115B,Bouwens2015ApJ...803...34B}). 
At higher redshifts, the detection of GNz11 by \cite{Oesch2016ApJ...819..129O} previously challenged the validity of these dust models at $z{>}10$ \citep{Mutch2016MNRAS.463.3556M}. This is evident in Fig. \ref{fig:glfs} with the predicted number density being barely consistent with the inferred value by GNz11 (but see the latest spectroscopic result from \citealt{Bunker2023arXiv230207256B} which resets GNz11 with a fainter magnitude and lower redshift). However, the latest JWST result for fainter galaxies from \cite{Donnan2023MNRAS.518.6011D} and \cite{Harikane2023ApJS..265....5H} suggests that the model prediction is consistent with observations up to $z{\sim}14$. Should the dust model fail at $z{>}10$ for the brightest galaxies (see e.g. \citealt{Ferrara2022arXiv220800720F,Markov2023arXiv230411178M}), the detection of GLz12 presents a serious challenge to our model as its luminosity function at $z{\sim}12$ {\it when ignoring dust attenuation} is still ${>}2$ times lower than GLz12 implies.

To explain the properties of these luminous galaxies, \citet[][]{Harikane2023ApJS..265....5H} considered modifying the initial mass function\footnote{Observational data shown in this work have been converted accordingly to match our \citet{Kroupa2001MNRAS.322..231K} IMF using the \textsc{astrodatapy} package (\url{https://github.com/qyx268/astrodatapy}).} (IMF) and incorporate a top-heavy, PopIII-dominated IMF to increase the intrinsic UV luminosity (see also recent theoretical work by \citealt{Haslbauer2022ApJ...939L..31H}, \citealt{Parashari2023arXiv230500999P}, \citealt{Shen2023arXiv230505679S}, \citealt{trinca2023exploring}, and \citealt{yung2023ultrahighredshift}). \meraxes is being upgraded to enable accurate modelling of PopIII star formation to address this possibility (Ventura et al. in prep.). In this work, we limit our exploration to the supernova feedback.

Supernova feedback is modelled as a thermal and kinetic source to inhibit gas collapse and star formation. Its efficiencies are tied to the maximum circular velocity ($V_{\rm max}$) of the host halo and increase in galaxies with shallower gravitational potentials \citep{Murray2005ApJ...618..569M,Guo2011MNRAS.413..101G}. While the energy coupling efficiency has no redshift dependence for a given $V_{\rm max}$, earlier results suggest larger mass-loading factors are of necessity to heat more gas in the early Universe \citep{Hopkins2014MNRAS.445..581H, Hirschmann2016MNRAS.461.1760H, Cora2018MNRAS.479....2C}. This again is based on matching relatively lower redshift observations (c.f. what we study in this work) and could fail at the early stages of the Epoch of Reionisation (EoR). Assuming no supernova feedback at all (\textit{no dust or fb} in Fig. \ref{fig:glf_z12}) leads to higher number densities even with respect to the upper limits from the cosmic-variance-free results of SuperBoRG at relatively lower redshifts \citep[][see also \citealt{Bagley2022arXiv220512980B}]{Leethochawalit2022arXiv220515388L}. We can further enhance the number density by increasing the star formation efficiencies\footnote{Our fiducial model (as well as \textit{no dust or fb}) takes 10x the dynamical time to deplete gas on the disc and form stars, roughly 1/20 of the Hubble time.} (i.e. \textit{maxSF}). These suggest that adjusting feedback or star-forming efficiencies at $z{>}10$ is needed to better model these new observations from JWST.

\subsection{JWST $z{\sim}16$ candidates are inconsistent with the standard model.}\label{subsub:z16}
The necessity of alternative galaxy-formation models has become increasingly pressing as observations have moved towards higher redshifts. The right panel of Fig. \ref{fig:glf_z12} highlights our fiducial predictions for the galaxy population at $z{\sim}16$, which presents very few galaxies that are brighter than -20 mag at these redshifts. However, the faintest reported $z{\sim}16$ candidate has a UV magnitude of ${-}20.4\pm0.2$. This implies that there are no analogues for any of the $z{\sim}16$ candidates in our fiducial catalogue. 

One might argue that these early JWST surveys are limited to a small effective volume and can be potentially biased by sample variance (see estimates by \citealt{yung2023ultrahighredshift}). However, our fiducial/standard galaxy-formation model extrapolates to a number density of ${\lesssim}10^{-12}{\rm Mpc}^{-3} {\rm mag}^{-1}$ at the magnitude of the brightest candidate ($M_{1600}{=}{-}21.6$). This implies that if these galaxies are spectroscopically confirmed to be at $z{\sim}16$, they will only exist in our fiducial outputs if the simulation volume covers the entire observable Universe. Given this, we argue that the existence of those $z{\sim}16$ candidates is inconsistent with our standard galaxy-formation model, which we emphasize again has a predicted galaxy population that is statistically representative of observations at relatively lower redshifts and/or luminosities.

At such high redshifts, dust is expected to have been produced in only trace amounts and therefore to not affect the UV luminosity function. However, we can still only reproduce $z{\sim}16$ galaxies with intrinsic UV magnitudes around $-21$ when feedback is assumed ineffective. In fact, to be consistent with the estimated number density for UV bright galaxies at $z{\sim}16$ \citep{Harikane2023ApJS..265....5H}, we need to effectively turn off supernova feedback and maximize star formation efficiency (see \textit{maxSF} in the right panel of Fig. \ref{fig:glf_z12}).
It is worth noting that when \citet[][see also \citealt{Donnan2023MNRAS.518.6011D} and \citealt{Naidu2022arXiv220802794N}]{Harikane2023ApJS..265....5H} was estimating the number density of $z{\sim}16$ bright galaxies, CEERS-93316 was considered as a candidate. 
However, a recent NIRSpec result has refuted this $z{\sim}16$ nature and determined it is in fact a low star-forming dusty galaxy at $z=4.9$ \citep{haro2023spectroscopic}. On the other hand, there are two additional $z{\sim}16$ candidates reported by \citet[][and studied further in the next section]{Atek2023MNRAS.519.1201A} which remain promising. Since they were found in a sub-field that \citet{Harikane2023ApJS..265....5H} explored, a revised number density that considers these two candidates are likely to be at a similar level as estimated by \citet{Harikane2023ApJS..265....5H}.

The detection of these extremely bright candidates at $z{\sim}16$ further illustrates that while feedback and regulated star formation are essential to galaxy formation across most cosmic time, this may not be the case at $z{\gtrsim}16$.

\section{JWST galaxies, observed and modelled}\label{sec:analogues}
In this section, the eight high-redshift JWST galaxies (see Table \ref{tab:galaxies}) are discussed in order of their redshifts and luminosities -- we start from intrinsically faint objects at relatively low redshifts, and then move onto bright ones found at much earlier times which also becomes increasingly challenging to study their analogues sometimes even with additional tuning of our model.

\subsection{JADES-GS-z13 \& JADES-GS-z12} \label{jadesz12}

\cite{Curtis-Lake2022arXiv221204568C} and \cite{Robertson2022arXiv221204480R} reported four spectroscopically confirmed galaxies at $z{>}10$. The two targets studied here, JADES-GS-z13 and JADES-GS-z12 of redshifts $z{=}13.30_{-0.07}^{+0.04}$ and $z{=}12.63_{-0.08}^{+0.24}$ respectively, come from an epoch even earlier than the previous record of high-redshift galaxies with spectroscopic confirmation, GNz11 (\citealt{Oesch2016ApJ...819..129O}, see also \citealt{Bunker2023arXiv230207256B} for an updated spectrum with NIRSpec). These galaxies are results from the JWST Advanced Deep Extragalactic Survey (JADES) that combines NIRCam and NIRSpec targeting at the GOODS (i.e., Great Observatories Origins Deep Survey) South (GS) field, reaching a $5\sigma$ magnitude limit of ${\sim}28.4$ mag for spectroscopy. When fitting the SEDs, \cite{Curtis-Lake2022arXiv221204568C} utilised the full spectra while \cite{Robertson2022arXiv221204480R} focused on the photometric data, leading to similar (but not identical) physical properties --  both JADES-GS-z13 and JADES-GS-z12 are quite small with an intrinsic UV magnitude fainter than -19 mag and a stellar mass of only ${\sim}10^8{\rm M}_\odot$. 

\subsubsection{Analogue selection}
In this work, we focus on the SEDs and inferred galaxy properties reported by \cite{Robertson2022arXiv221204480R} when identifying analogues within our simulation. In particular, we look for modelled galaxies that have an SED consistent with the measurement by requiring the magnitude in bands F200W, F277W, F356W and F444W (i.e., the ones above the Lyman-$\alpha$ break) to be within 2$\sigma$ of the observational uncertainties\footnote{Throughout this paper, a 10\% error floor is additionally considered in the observed photometry for all targets to account for potentially underestimated systematics (see e.g. \citealt{Naidu2022ApJ...940L..14N}). This includes the two $z{\sim}16$ candidates found in lensing fields (see Section \ref{subsection:smacz16ab}) before errors of the lensing model are further added. In addition, the negative 2$\sigma$ threshold of non-detection is reset to zero during analogue selection.} while luminosities also have to be under the 2$\sigma$ threshold for bands of non-detection (i.e. F090W, F115W and F150W). We further apply a prior on the redshift range (i.e., $z\in[10.8,16.9]$ over 31 snapshots) to avoid low-redshift contamination and speed up the selection process. 

It is worth highlighting an implicit prior built into our analogue selection as a result of the evolving luminosity function -- within a cosmological simulation box, there are more galaxies with fainter magnitudes or lower redshifts (see Fig. \ref{fig:glfs}). Therefore, when marginalising the analogue sample distribution onto luminosity-vs-redshift, galaxies with low luminosities and redshifts are dominant. This often leads to lower values of these two properties (and other properties sharing a degeneracy) compared to observational results that do not impose such a prior.

\begin{figure*}
	\centering
	\includegraphics[width=1.0\textwidth]{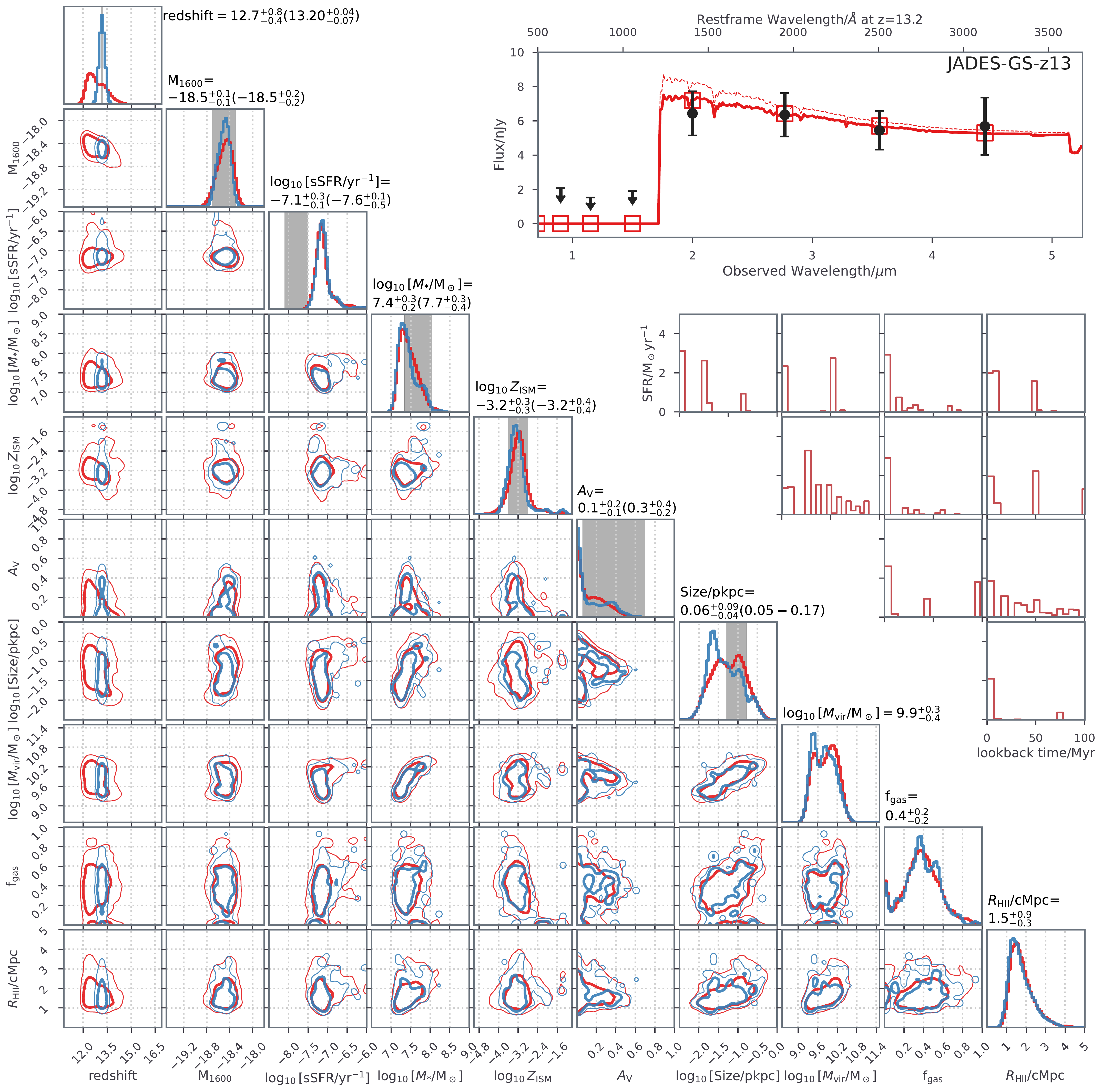}\\\vspace*{-3mm}
	\caption{JADES-GS-z13 analogues. \textit{Lower-left corner plot:} marginalized galaxy property distributions of the model analogues with the redshift prior set to be between z=10.8 and 16.9 (red) as well as based on the $2\sigma$ uncertainties of the spectroscopic result (blue, \citealt{Curtis-Lake2022arXiv221204568C}), respectively. Note that in the 1D distributions, the vertical axes present number densities in linear scale with the total integral normalized to 1 for both. From left to right or top to bottom, these are redshift, intrinsic UV magnitude, star formation rate averaged over 30Myr and normalised by stellar mass (sSFR), stellar mass, metallicity, dust extinction, galaxy size, halo virial mass, the fraction of gas that is accessible to star formation, and the radius of surrounding ionized bubble. Indicated on the top of each 1D distribution are the median value and $1\sigma$ uncertainties ([16,84] percentiles) based on the larger redshift prior. For comparison, estimates from \citet{Robertson2022arXiv221204480R} are shown as the grey shaded regions and inside the parentheses. \textit{Top right panel:} modelled SED and spectra for an example analogue with thick solid and thin dashed lines indicating whether dust attenuation is considered. The nominated SEDs and $2\sigma$ uncertainties from \citet[][using \textsc{forcepho}]{Robertson2022arXiv221204480R} are shown with black circles while upper limits are presented as $5\sigma$.
		\textit{Central-right corner plot:} Star formation rate in the past 100Myr for 10 randomly selected analogues, to illustrate the bursty star formation nature of these low-mass galaxies in our simulation as a reason for the inferred high sSFR.}
	\label{fig:JADEz13}
\end{figure*}

\begin{figure*}
	\centering
	\includegraphics[width=\textwidth]{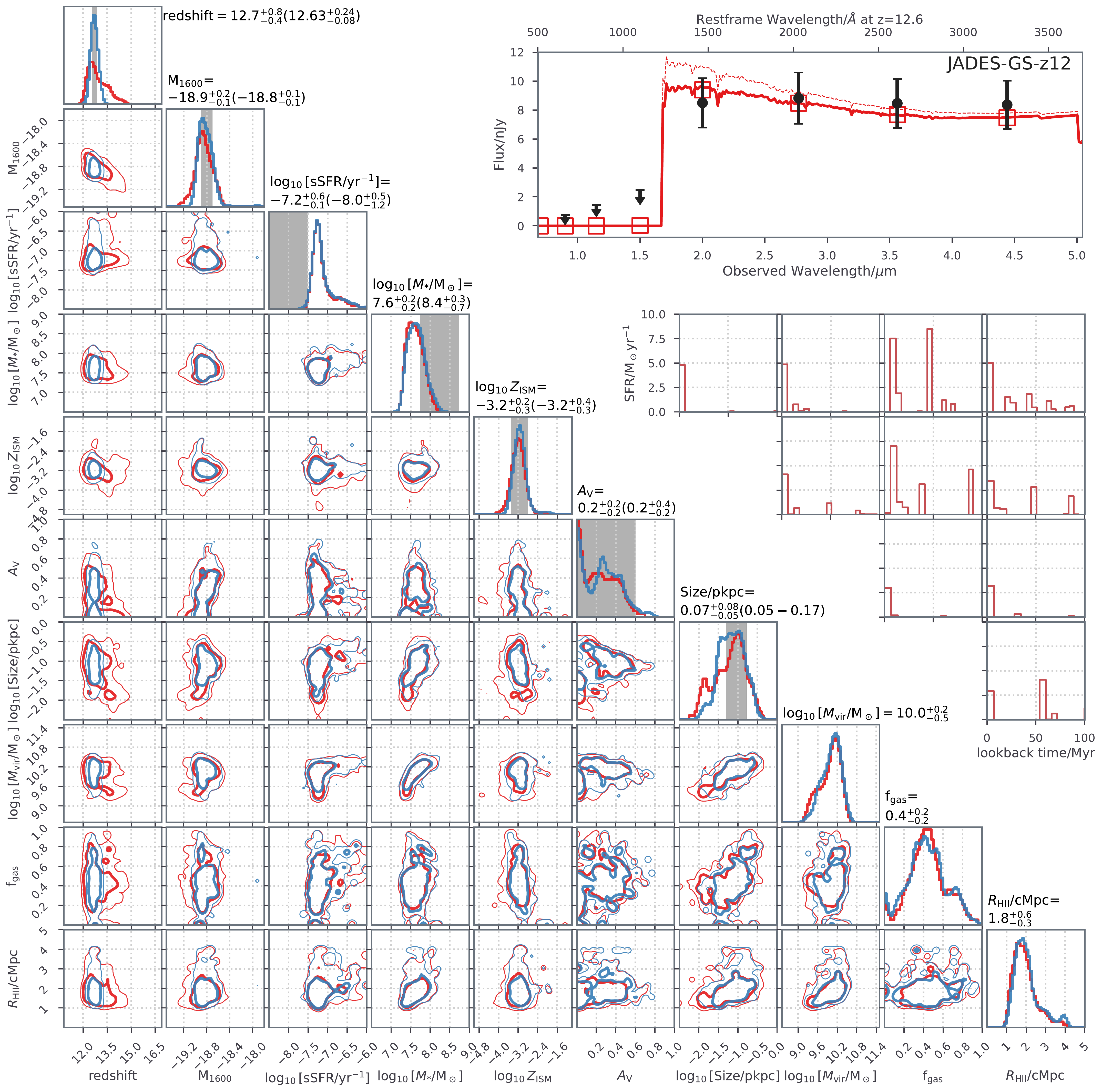}\\\vspace*{-2mm}
	\caption{Similar to Fig. \ref{fig:JADEz13} but for JADES-GS-z12 and its analogues.}
	\label{fig:JADEz12} 
\end{figure*}

Our selection leads to 1296 and 397 analogues in our fiducial output for JADES-GS-z13 and JADES-GS-z12, respectively, whose properties are summarized in Figs. \ref{fig:JADEz13} and \ref{fig:JADEz12}.


\subsubsection{Galaxy properties}
We find both observed SEDs to be consistent with modelled star-forming galaxies at $z~{\sim}12.6$ or ${\sim}13.2$, as is evident from the two modes in the redshift distribution. With minor differences, the high redshift mode also indicates higher luminosities, lower metallicities and less dust extinction. However, the overall distribution does not alter significantly after applying the redshift prior inferred by the spectral break at Lyman-$\alpha$ \citep[][i.e. comparing the blue and red distributions]{Curtis-Lake2022arXiv221204568C} with the predicted galaxy properties comparable to estimates from \citet{Robertson2022arXiv221204480R} -- both JADES-GS-z13 and JADES-GS-z12 analogues have an intrinsic UV magnitude around $M_{1600}=-18.5$, a stellar mass less than $10^8{\rm M}_\odot$ and a size of only ${\sim}60$ physical parsecs with very low metallicities and suffering little dust attenuation. 

The major difference between our result and, for instance, \citet{Robertson2022arXiv221204480R} comes from the star formation history (SFH). Already flexible, the SFH considered by \citet[][see also other observational studies mentioned in this work]{Robertson2022arXiv221204480R} during the SED fitting includes 6 snapshots between $z{\sim}12$ and 20 in order to capture the burstiness of high-redshift star formation. On the other hand, there are 35 snapshots in our simulation between these redshifts and, with such a high cadence, we are able to accurately simulate the SFH in the presence of time-resolved feedback (massive stars can take much longer to become supernovae than the dynamical timescale of the galactic disc; \citealt{Mutch2016MNRAS.463.3556M}). In addition, our model does not impose the bursty-continuity prior as in \citet{Robertson2022arXiv221204480R}, but rather requires galaxies to accumulate enough cold and dense gas before being able to form stars. For instance, Figs. \ref{fig:JADEz13} and \ref{fig:JADEz12} show that JADES-GS-z13 and JADES-GS-z12 are likely hosted by halos of ${\sim}10^{10}{\rm M}_\odot$ and, with around 40 per cent of their gas accessible to star formation ($f_{\rm gas}$)\footnote{\textit{Meraxes} reserves a so-called ejected gas reservoir for each modelled galaxy, as a response to supernova feedback. Gas in this reservoir is considered to have a cooling timescale much longer than the Hubble time and therefore does not contribute to star formation.}, these galaxies might have 5 times more (in mass) star-forming gas than their stellar components. Also because our modelled galaxies have to reach this critical mass before forming stars, the analogues possess a much burstier SFH than \citet{Robertson2022arXiv221204480R}, leading to higher recent star formation rates (SFRs averaged over the past 30Myr; and specific SFR) and/or lower integrated stellar masses.

Finally, our model also predicts that (assuming all galaxies have a UV ionizing escape fraction of 0.15) JADES-GS-z13 and JADES-GS-z12 are likely located in ionized bubbles of ${\lesssim}2$ cMpc in radius\footnote{500 sight-lines towards the target are randomly drawn and the distance between the galaxy and the nearest cell on each sight-line that has a neutral fraction no less than 90 per cent is measured. 50, 16, and 84 percentiles are then calculated with the median nominated as the \textsc{Hii} bubble radius.}. However, in rare cases where the analogue coexists with a more massive neighbour, it may have a much larger ionized bubble of up to ${\sim}4$ cMpc in radius (see also \citealt{Qin2022MNRAS.510.3858Q,Whitler2023arXiv230516670W}).

\subsection{S5-z12-1}\label{S5-z12-1}

\begin{figure*}
	\centering
	\includegraphics[width=\textwidth]{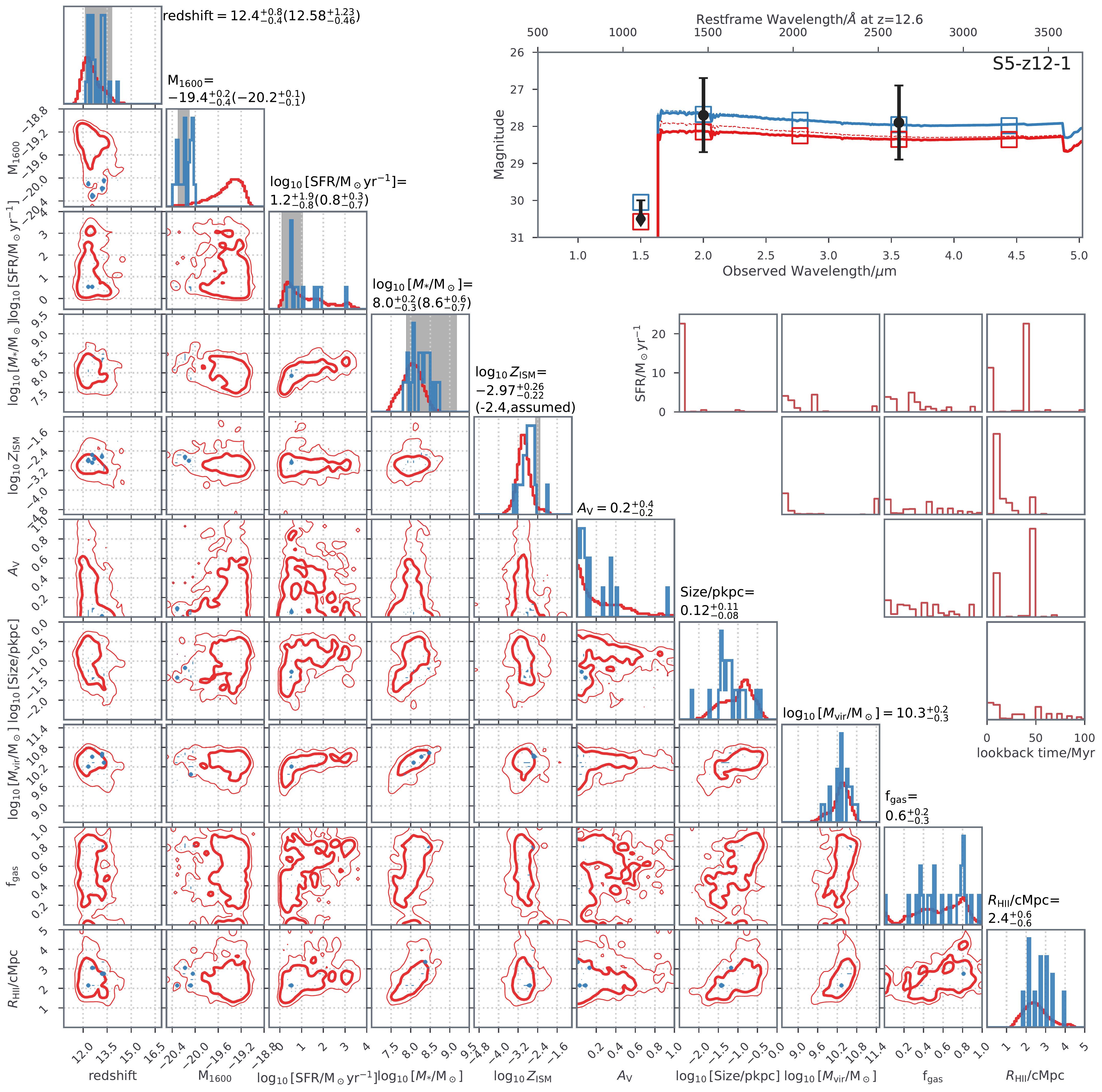}\\
	\caption{Similar to Figs. \ref{fig:JADEz13} and \ref{fig:JADEz12} but for S5-z12-1 and its analogues. Note the upper limit in band F150W is $2\sigma$ and the blue colour now indicates a small number of analogues having ${\rm M}_{1600}$ consistent with the $2\sigma$ uncertainties of the inferred luminosities from \citet{Harikane2023ApJS..265....5H}. Since the randomly selected analogue spectrum represents a low-luminosity galaxy, the top-left panel also highlights a second analogue with an intrinsic magnitude comparable to the observation. To ease comparison we also replace the sSFR panels with SFR, which is now an averaged over 50Myr following \citet{Harikane2023ApJS..265....5H}.}
	\label{fig:S5-z12-1} 
\end{figure*}

\cite{Harikane2023ApJS..265....5H} analyzed multiple NIRCam fields, producing a comprehensive study of high-redshift JWST galaxies. Their final sample consists of 10 $z{\gtrsim}12$ candidates, mostly from CEERS (i.e., Cosmic Evolution Early Release Science led by \citealt{Finkelstein2022ApJ...940L..55F}) as well as GLASS (an ERS program led by \citealt{Treu2022ApJ...935..110T}), SMACS~J0723.3-7327 (hereafter SMACS; a $z{=}0.39$ galaxy lensing cluster that has previously been searched for high-redshift candidates, e.g. \citealt{Coe2019ApJ...884...85C}), and Stephan's Quintet (a group of five local galaxies). 

As the total effective area probed by these fields amounts to ${\sim}90\ {\rm deg}^2$, \citet{Harikane2023ApJS..265....5H} estimate the galaxy UV luminosity function out to $z{\sim}16$. Figure \ref{fig:glfs} shows that our fiducial model is consistent with their result at the faint end, echoing the large analogue sample that we have successfully identified for faint galaxies such as the two JADES-GS galaxies (see also \citealt{McCaffrey2023arXiv230413755M}) in Section \ref{jadesz12}. However, a significantly lower number density for bright galaxies is also seen from our simulation result, indicating that the identification of analogues is increasingly challenging for more luminous candidates.

Here, we focus on candidates from \citet{Harikane2023ApJS..265....5H} that have a UV magnitude brighter than ${\sim}$-20. Among these, we first present analogues for the faintest\footnote{In fact, there is another object, CR2-z12-1, which is considered as a fainter $z{\sim}12$ galaxy with an intrinsic UV magnitude of ${-19.9}{\pm}0.1$ by \citet{Harikane2023ApJS..265....5H}. However, \citet{Finkelstein2022ApJ...940L..55F} considers the same object, which they call Maisie's Galaxy, as -20.3 mag. We return to this candidate in Section \ref{CEERSz14}. Furthermore, S5-z12-1 was initially considered at $z{=}13.72^{+0.86}_{-1.92}$. Using the earlier photometric results from the pre-print version, we reached a similar conclusion and saw lower intrinsic luminosities w.r.t. the observation.} target in this section, S5-z12-1, which is a $z{=}12.58^{+1.23}_{-0.46}$ galaxy with an intrinsic UV magnitude of ${-}20.2{\pm}0.1$ found in Stephan's Quintet.

\subsubsection{Analogue selection}
Photometry from only 3 filters was reported by \citet{Harikane2023ApJS..265....5H} for S5-z12-1. For analogue selection, we require $2\sigma$ consistency in band F200W and F356W between the forward modelling and the observational data; while the inferred magnitude in the non-detection band, F150W, must be lower than the $2\sigma$ threshold. We use the same 31 snapshots as before and focus on $z{\sim}11$--17. We end up with a surprisingly large number of 377 analogues, among which, however, only 22 have {\it an intrinsic UV magnitude} $2\sigma$ consistent with the estimate of \citet{Harikane2023ApJS..265....5H}. We show a similar summary plot as before for S5-z12-1 in Fig. \ref{fig:S5-z12-1} but with the blue colour highlighting analogues with the high UV luminosity inferred by the observation.

\subsubsection{Galaxy properties}
Figure \ref{fig:S5-z12-1} shows that our inferred properties are mostly consistent with those of \citet[][compare 1D distributions with grey shaded regions]{Harikane2023ApJS..265....5H}, except for the intrinsic UV magnitude -- our model suggests S5-z12-1 is a more massive, larger and brighter galaxy than the two confirmed $z{\sim}12$ galaxies but still with a low metallicity and minor dust attenuation. It has adequate gas available for star formation thanks to its $10^{10}$--$10^{10.5}{\rm M}_\odot$ halo mass. This indicates that S5-z12-1 is likely located in a more over-dense region and hence surrounded by a larger ionized bubble of ${\sim}2.5$ cMpc radius.

The significantly lower luminosity (than \citealt{Harikane2023ApJS..265....5H} inferred) we obtain for S5-z12-1 is likely caused by our predicted $z{\sim}12$ UV luminosity function dropping to a level lack of statistical meaning at $M_{1600}$ around $-20.2$. Therefore, when seeking analogues in our modelled cosmological lightcone where galaxy number evolves with not only redshift but also luminosity, we preferentially find galaxies on the higher magnitude end of the observed photometry uncertainties (see e.g. the red line in the SED panel of Fig. \ref{fig:S5-z12-1}). When focusing on the 22 analogues that have the intrinsic UV magnitude between $-20.4$ and $-20$, the inferred properties become more consistent with the observational results. This includes SFR and stellar mass, which were shown in Section \ref{jadesz12} to be systematically different from observations that assume the continuity prior for SFH. This improved consistency is likely due to the slightly larger masses of S5-z12-1 analogues, which facilitate a smoother (i.e., continuous) SFH in our model.

In the next two sections, we discuss even brighter $z{\sim}12$ targets. For them, because of the low predicted number density as seen by our fiducial model, we will no longer obtain a statistically meaningful analogue sample. Rather, the few analogues we present below can only be interpreted as \textit{possible} solutions for these JWST high-redshift galaxies. This includes their evolutionary paths and contribution to the local ionization that we will present in more detail.

\subsection{GLz12} \label{gnlz12}

\begin{figure*}
	\centering
	\includegraphics[width=.87\textwidth]{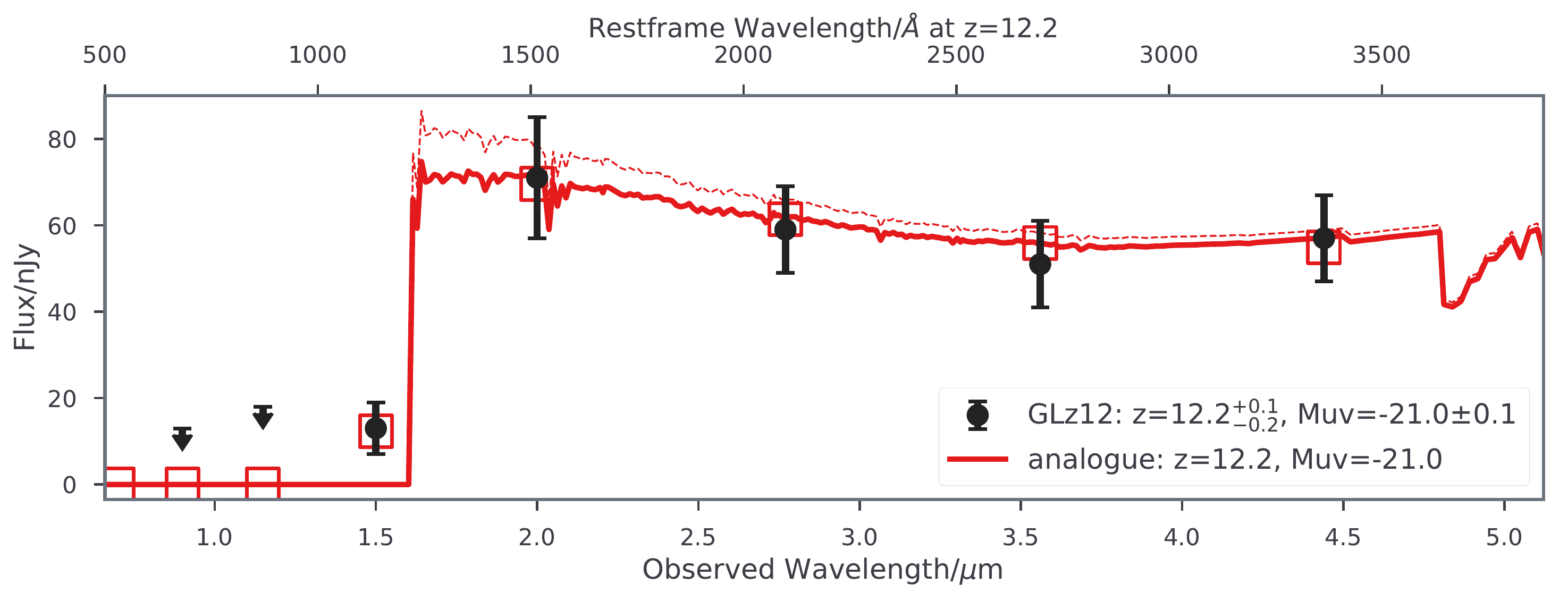}\\
	\begin{minipage}{0.41\linewidth}
		\includegraphics[width=\linewidth]{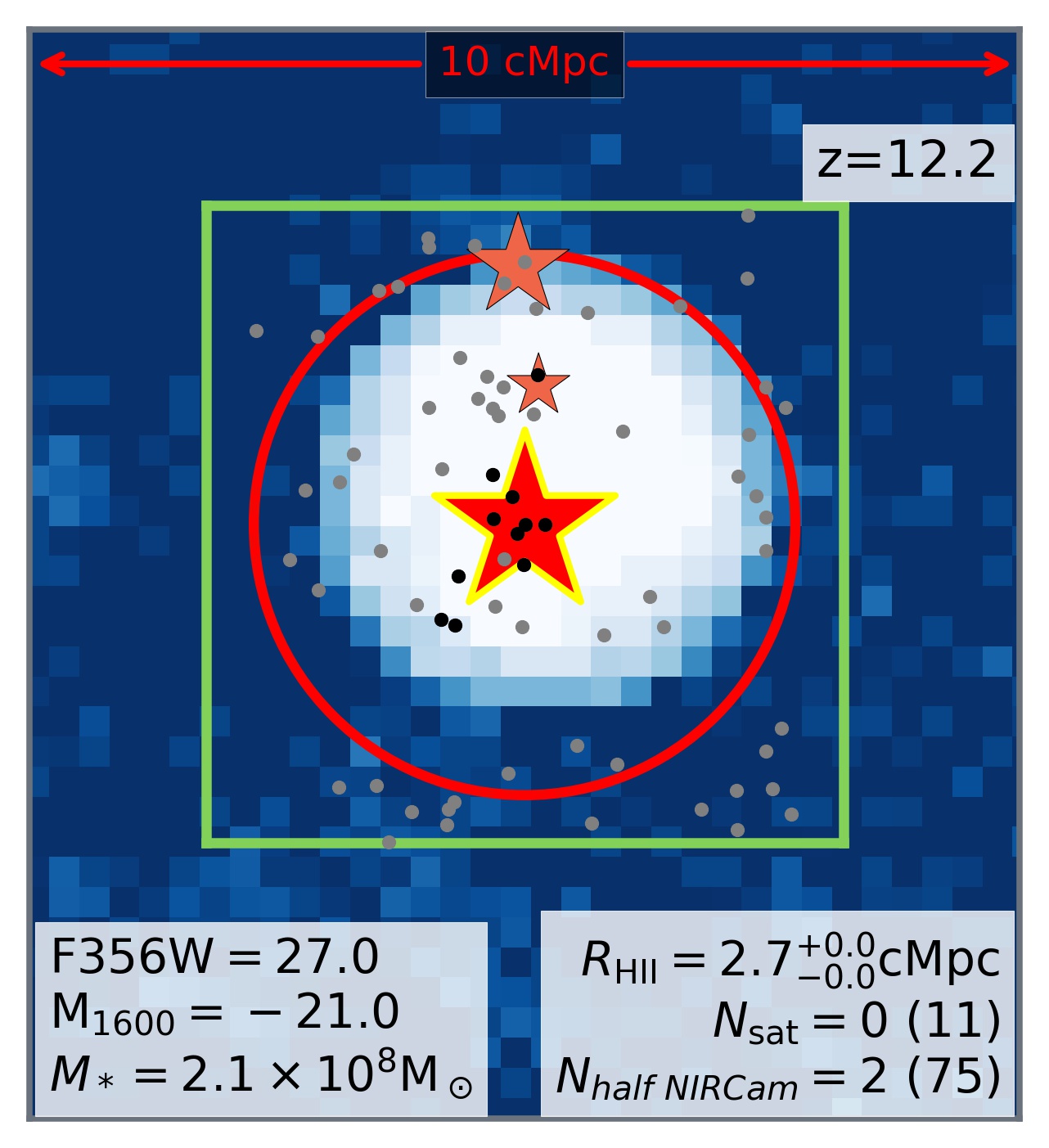}
	\end{minipage}
	\begin{minipage}{0.217\linewidth}
		\begin{minipage}{\linewidth}
			\includegraphics[width=\linewidth]{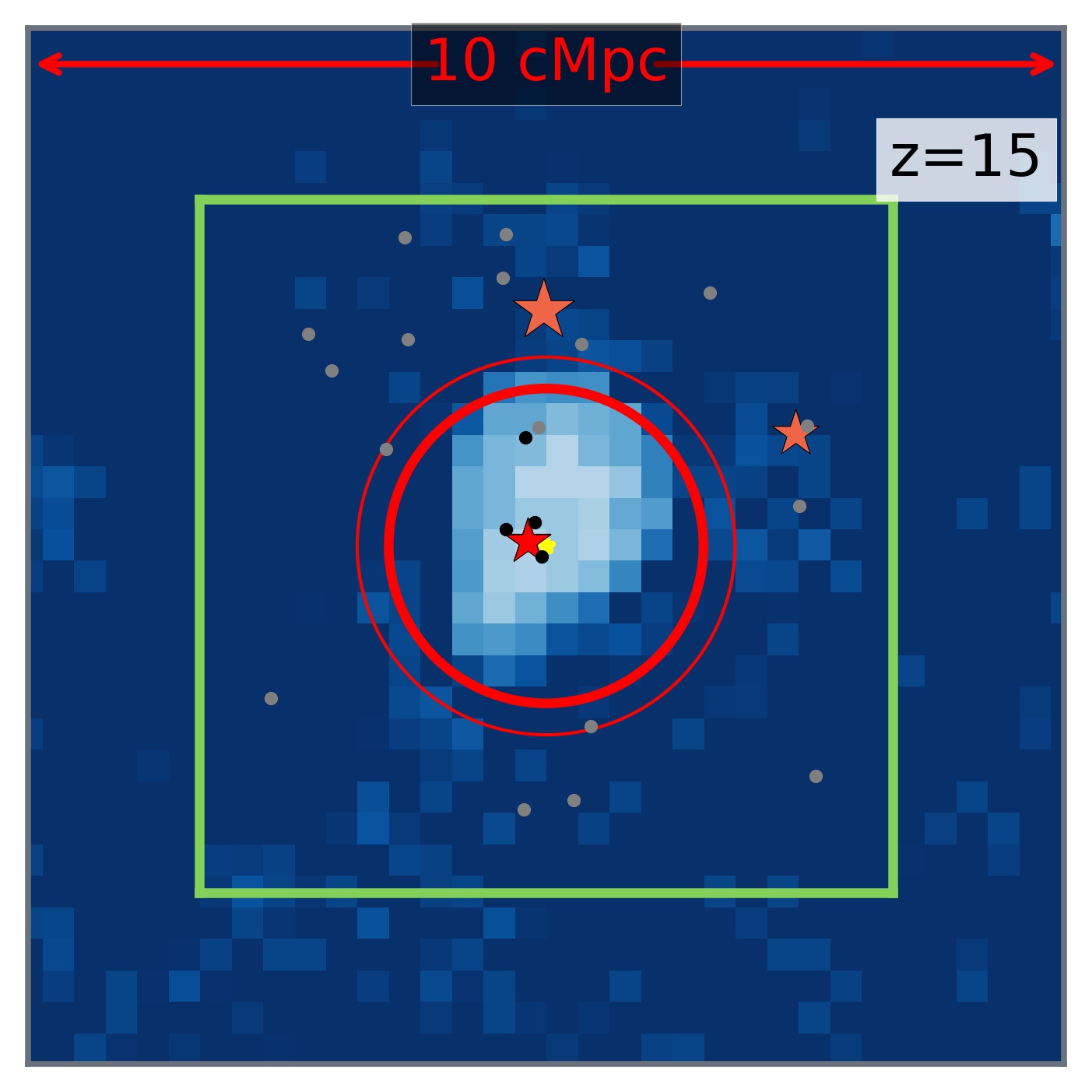}
		\end{minipage}\\
		\begin{minipage}{\linewidth}
			\includegraphics[width=\linewidth]{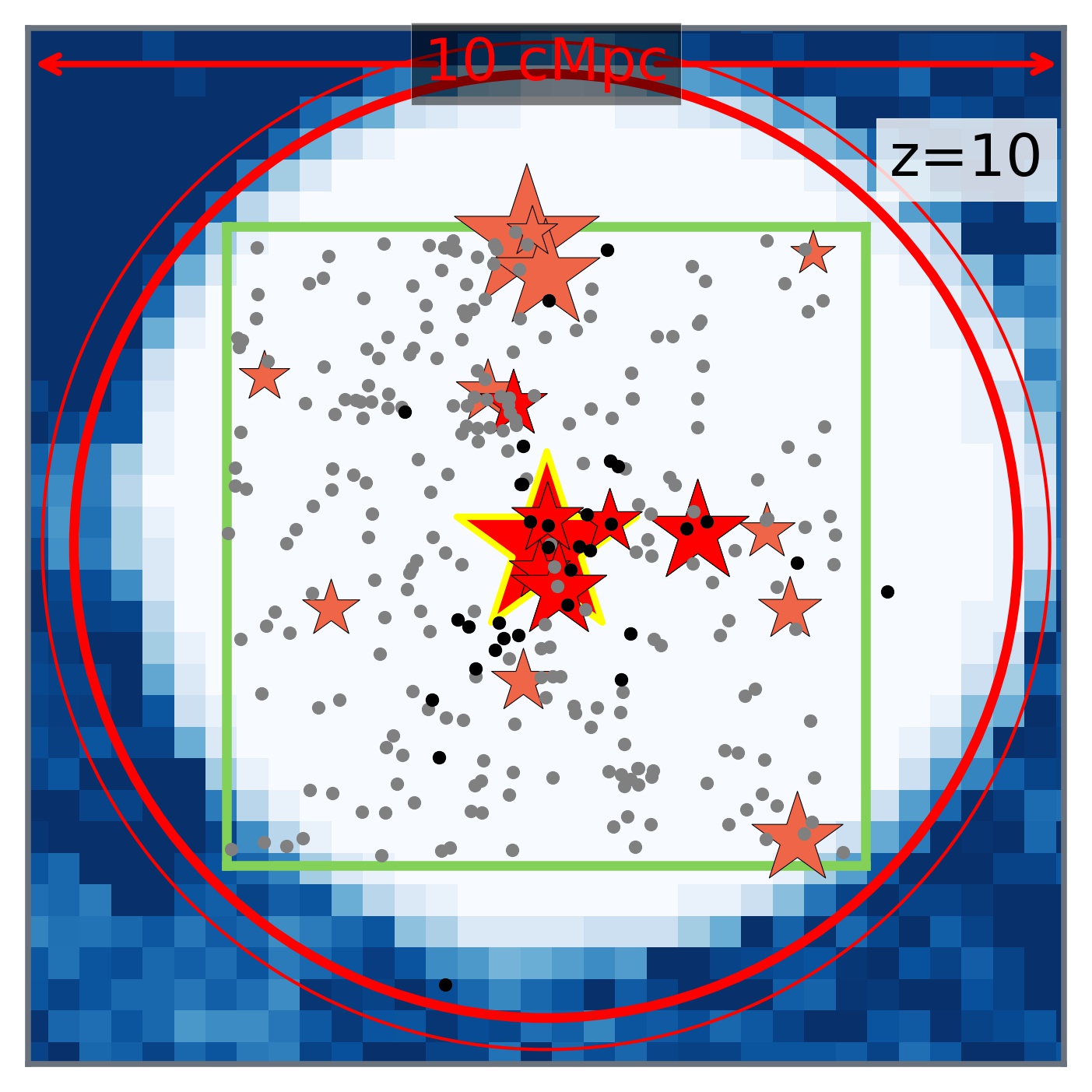}
		\end{minipage}
	\end{minipage}
	\begin{minipage}{0.36\linewidth}
		\begin{minipage}{\linewidth}
			\includegraphics[width=\linewidth]{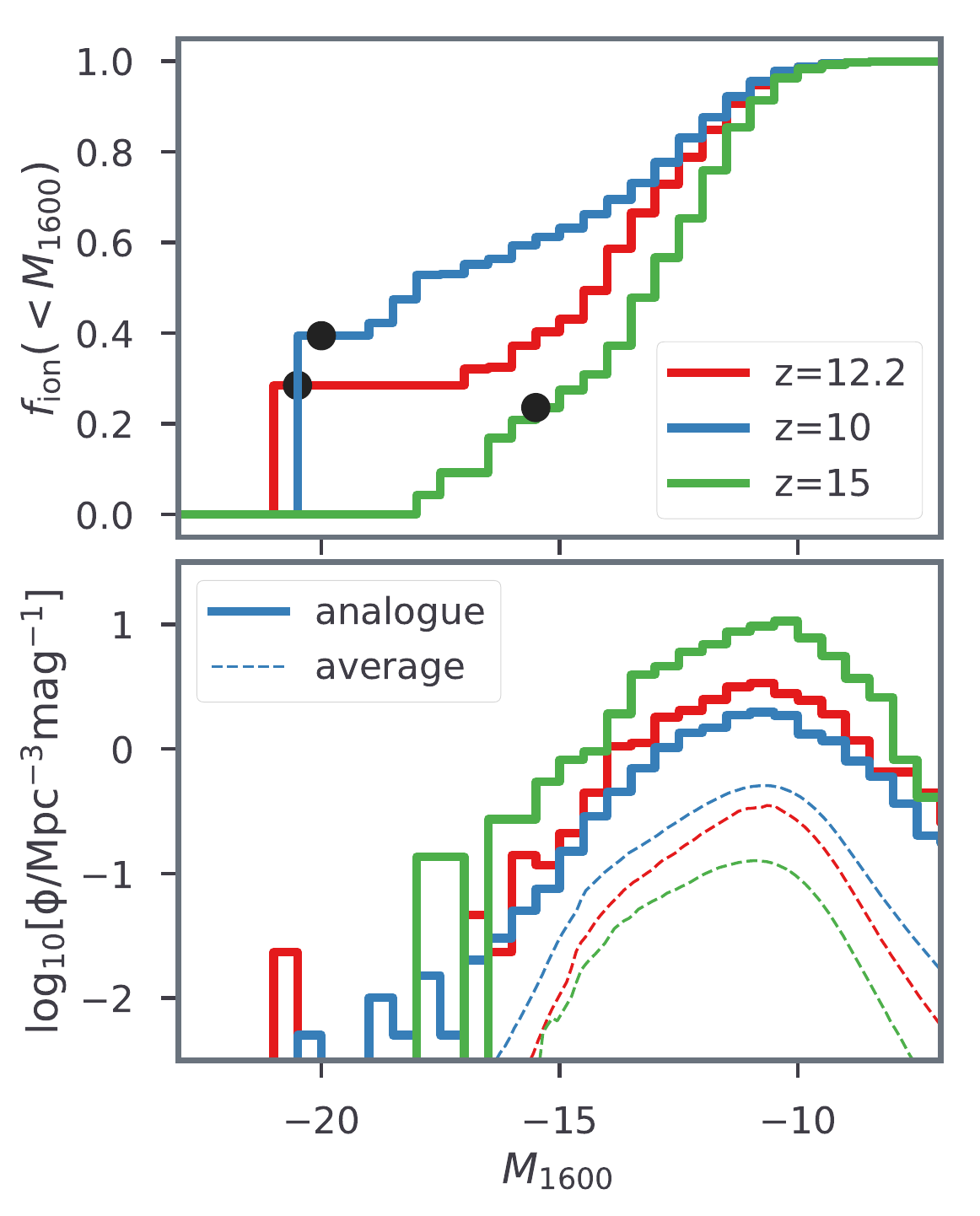}
		\end{minipage}
	\end{minipage}
	\caption{GLz12 analogue. \textit{Top panel:} modelled SED and spectra are shown in colour	while the nominated observed SED and the $2\sigma$ uncertainties or upper limits from \citet{Naidu2022ApJ...940L..14N} are presented in black. \textit{Lower-left panel:} local environment at the observed redshift with background indicating \textsc{Hi} ionization (projection depth and colour follow Fig. \ref{fig:lightcone}). The apparent and intrinsic UV magnitudes are listed in the lower-left corner for the analogue followed by its stellar mass. Presented in the lower-right corners are the radii of the \textsc{Hii} bubble (with the mean indicated by thick red circles and thin ones for the $1\sigma$ uncertainties), the number of neighbouring bright galaxies (F356W${<}30.7$ mag) within the mean \textsc{Hii} bubble (indicated by filled stars) and that (indicated by the transparent stars if not inside the bubble) within half of {\it JWST} NIRCam FoV ($2.2'{\times}2.2'$; 2D projection with $\Delta z{\sim}{\pm}0.5$, see the green squares). Numbers in the brackets indicate counting down to F356W${=}32.7$ mag (with dark and light dots indicating faint galaxies within the bubble or within the FoV). These two magnitude thresholds are motivated by upcoming {\it JWST} deep and lensing fields, respectively. \textit{Lower-middle panels:} local environment at earlier and later times to visualize the evolution around the analogue. \textit{Lower-right panels:} the normalized cumulative number of UV ionizing photons and number density as a function of UV magnitudes for galaxies within the ionized region. The UV magnitude of GLz12 analogue as well as its progenitor and descendent is indicated by the circles.
	}
	\label{fig:glf_z12_analogues_sed}
\end{figure*}

\citet{Naidu2022ApJ...940L..14N} was among the first (see also \citealt{Castellano2022ApJ...938L..15C}) to report $z{\gtrsim}10$ galaxies soon after JWST images became publicly available\footnote{Throughout this paper, a few footnotes present a brief summary of our analogue study when earlier observational results including the SEDs were utilized. This is to demonstrate the stochasticity of analogue studies as minor changes in the SED can lead to different identifications. However, while the comparison addresses the sampling issue when studying these bright galaxies, the conclusion from using different versions of SED remains qualitatively consistent. For instance, GLz12 was previously considered at $z{=}13.1^{+0.8}_{-0.7}$ by \citet{Naidu2022ApJ...940L..14N} with the initial NIRCam calibration. Using the earlier SED from its pre-print, we identified 2 analogues showing qualitatively similar evolution and environment as the analogue presented here.}. This includes GLz12 from GLASS, a $z{=}12.2^{+0.1}_{-0.2}$ target that we discuss in this subsection. Its inferred properties include an average SFR of ${\sim}10{\rm M}_\odot {\rm yr}^{-1}$, a stellar mass around $10^{9}{\rm M}_\odot$, an intrinsic UV magnitude of $-21$, a low dust extinction of $A_{v}{\sim}0.3$ and an effective radius around 0.5 pkpc (see also \citealt{Ono2022arXiv220813582O}). This object was also identified by other independent groups \citep{Donnan2023MNRAS.518.6011D,Harikane2023ApJS..265....5H,Santini2023ApJ...942L..27S} who inferred somewhat different physical properties. In this work, we adopt the reported values from \cite{Naidu2022ApJ...940L..14N}.

\subsubsection{Analogue selection}

We use the reported photometry from \citet[][their table 2]{Naidu2022ApJ...940L..14N}, update the 1$\sigma$ uncertainties with a 10\% error floor, and seek galaxies in the 31 consecutive output snapshots between $z{=}10.8$ and 16.9 that have magnitudes consistent with observations -- we require the modelled SED in band F200W, F277W, F356W and F444W (all above the Lyman-$\alpha$ break) to be within 2$\sigma$ of the observational uncertainties while the flux also has to be below the 2$\sigma$ threshold for bands of non-detection (i.e. F090W, F115W and F150W).

We identify only one analogue, which is found at the nominated photometric redshift of GLz12 using \textsc{easy} \citep{Brammer2008ApJ...686.1503B}, i.e. $z{=}$12.2, with the same inferred intrinsic UV magnitudes of $-21.0$ mag. 
The top panel of Fig. \ref{fig:glf_z12_analogues_sed} shows the spectrum and SED of this analogue. As with the two spectroscopically confirmed galaxies, it also shows the spectral features of typical star-forming galaxies at high redshift with a UV slope of ${\lesssim}{-}2$ \citep{Bouwens2014ApJ...793..115B} and negligible dust attenuation.

\subsubsection{Local environment}
The lower panels of Fig. \ref{fig:glf_z12_analogues_sed} present the local environment of our single analogue of GLz12 -- from left to right, we illustrate the local ionization at $z=12.2$, 15 and 10 as well as the distribution of galaxies within the \textsc{Hii} bubble as in the normalized cumulative number of UV ionizing photons and the number density of galaxies as a function of luminosity. With a total stellar mass of $10^{8.3}{\rm M}_\odot$, this analogue has ionized its surrounding IGM to a radius of 2.7 cMpc, larger than the majority of analogues presented in Sections \ref{jadesz12} and \ref{S5-z12-1}. However, despite being the most massive galaxy in its ionized region and at least 4 magnitudes brighter than all the other neighbouring galaxies in the \textsc{Hii} bubble, this analogue only contributes ${\sim}30$\% of the local UV ionizing photons. Such a high and early local ionization is only possible when a significant portion (i.e. 15 per cent in the fiducial model) of ionizing photons manage to escape from the numerous low-mass galaxies that are as faint as $M_{1600}{\sim}$ $-15$ to $-12$ mag.

The analogue is also in an over-dense region of the universe. When counting the number of galaxies within the \textsc{Hii} region, the density is an order of magnitude higher than the average field inferred from the UV luminosity function (c.f. Fig. \ref{fig:glfs}). For instance, follow-up of GLz12 with deeper JWST observations (e.g. WDEEP led by \citealt{Finkelstein2021jwst.prop.2079F} aims to reach a magnitude limit of F356W=30.7) is expected to uncover another two neighbouring galaxies within 2.2'$\times$2.2' ($\Delta z{\sim}\pm0.5$ is considered). This is shown in Fig. \ref{fig:glf_z12_analogues_sed} (see also recent follow-ups on bright Lyman-$\alpha$ emitting galaxies at relatively lower redshifts, e.g. \citealt{Leonova2022MNRAS.515.5790L,Tacchella2023arXiv230207234T,Witten2023arXiv230316225W,Whitler2023arXiv230516670W}). However, among these neighbours, none sits inside the ionized bubble of the GLz12 analogue since reionization is still at its infant stage at these high redshifts and the ionized regions correspond to sizes of only $\Delta z\sim{\pm}0.01$.

When probing the progenitors and descendants of GLz12, our model suggests that it is among the first galaxies that start reionization and remains highly efficient in forming stars and contributing UV ionizing photons across cosmic times. At earlier redshifts such as $z\sim15$, the analogue sits in a non-spherical ionized region of around $1.5$ to 1.8 cMpc in radius, which is already more than half the size it will grow to by $z{=}12.2$. Its intrinsic UV magnitude is fainter than -16 mag. Therefore, unlike GLz12 at $z{=}12.2$, the progenitor is instead ${\sim}3$ magnitudes fainter than the brightest in the region and has a very minor contribution to reionization at these early times. On the other hand, its dominance grows at lower redshifts. For instance, with a stellar mass exceeding $10^9 {\rm M}_\odot$ at $z{=}10$, this descendant alone contributes 40 per cent in the local ionizing budget, expanding its \textsc{Hii} territory to nearly 5 cMpc in radius. From the lower-right panels of Fig. \ref{fig:glf_z12_analogues_sed},  we also find the ionized region is increasingly biased towards higher redshifts with fainter galaxies becoming relatively more significant to reionization.

\subsubsection{Formation history \& subsequent evolution}

\begin{figure}
	\centering
	\includegraphics[width=\columnwidth]{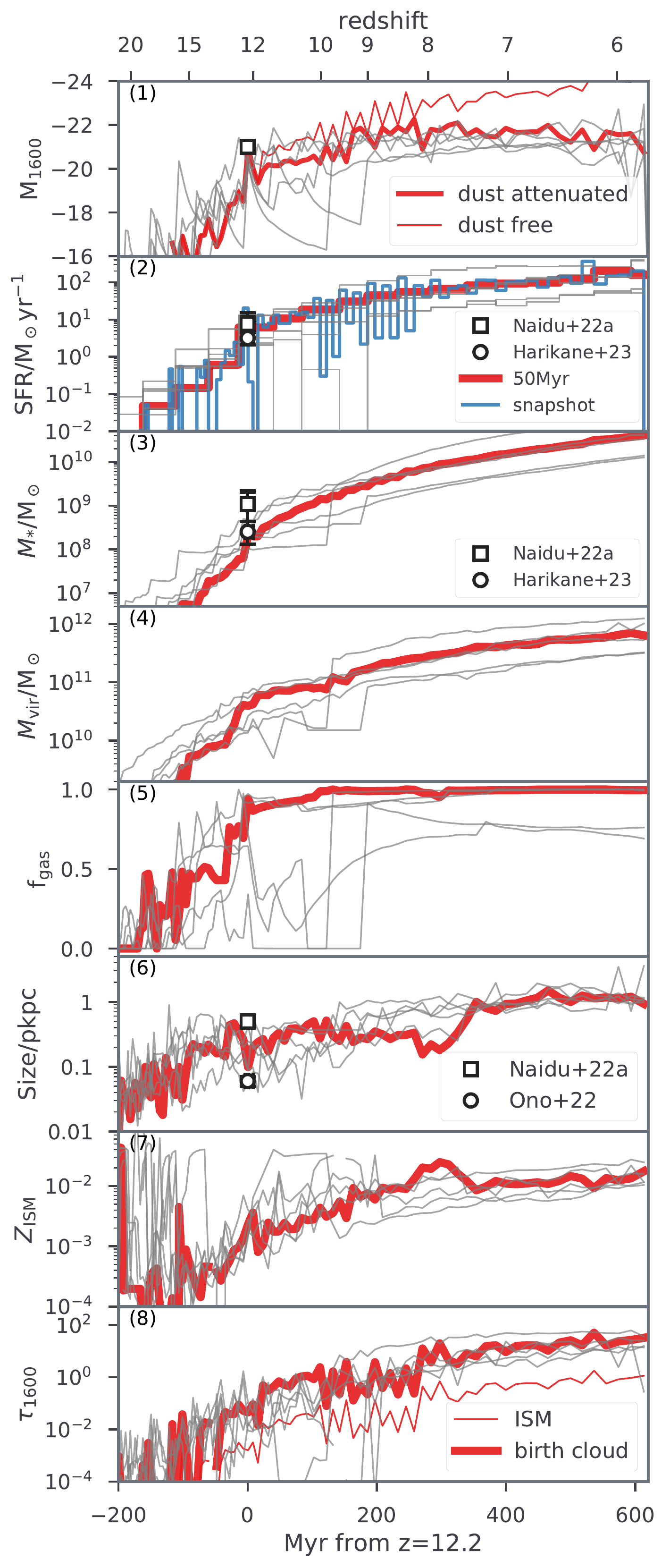}\\\vspace*{-2mm}
	\caption{GLz12 analogue history as in (1) the intrinsic UV magnitude (thick and thin lines consider or exclude dust attenuation); (2) SFR averaged over each snapshot (blue) or ${\sim}50$Myr following \citet[][red]{Naidu2022ApJ...940L..14N}; (3) stellar mass; (4) halo mass; (5) fraction of gas accessible to star formation; (6) galaxy half-light radius; (7) ISM metallicity; and (8) optical depths for photons with a wavelength of 1600{\AA} (thick and thin lines for $\tau_{1600}$ in the birth cloud of the emitting stars or in the ISM). GLz12 properties estimated by \citet{Naidu2022ApJ...940L..14N}, \citet{Harikane2023ApJS..265....5H} and \citet{Ono2022arXiv220813582O} are also indicated in corresponding panels. For comparison, thin grey lines illustrate property histories for the next five brightest galaxies (only dust attenuated ${\rm M}_{1600}$ and $\tau_{1600}$ in the birth cloud are shown to not crowd the plot).
	}
	\label{fig:glf_z12_analogues_histories}
\end{figure}

We next look at the possible evolutionary path of GLz12 using Fig. \ref{fig:glf_z12_analogues_histories}, in which the history of our analogue in terms of its UV magnitude, SFR, stellar mass, halo mass, gas content, size, metallicity and optical depth to UV non-ionizing photons is presented. The inferred properties using \textsc{Prospector} \citep{Leja_2017} reported in \cite{Naidu2022ApJ...940L..14N} as well as estimates from \citet{Harikane2023ApJS..265....5H} and \citet{Ono2022arXiv220813582O} are also shown in the figure for comparison. 

We see that the GLz12 analogue grows its UV luminosities very rapidly at $z{\ge}12$ with the analogue showing a 100$\times$ increment from $z{\sim}15$ to 12 which is less than $100$ Myr. When averaged over 50 Myr, this analogue possesses a steady growth of SFR in the past and reaches ${\sim}10{\rm M}_\odot {\rm yr}^{-1}$ at $z{\sim}12.2$, consistent with the observational results. On the other hand, the snapshot-averaged SFR presents a bursty SFH, similar to analogues of less massive galaxies discussed in the previous two sections. However, due to its relatively large dark matter component which has created a deeper gravitational potential, a greater amount of gas has been accreted by the GLz12 analogue to fuel star formation for a longer period of cosmic time. Therefore, its SFH is less bursty compared to the low-mass counterparts such as JADES-GS-z12 and JADES-GS-z13 (c.f. Figs. \ref{fig:JADEz13} and \ref{fig:JADEz12}). Our model predicts the GLz12 analogue has a stellar mass of only $2{\times}10^{8} {\rm M}_\odot$, which is 5 times lower than \citet{Naidu2022ApJ...940L..14N} but more consistent with \citet{Harikane2023ApJS..265....5H}. These two studies adopt the same continuity prior for the SFH, highlighting the potentially underestimated systematics in these ERS results which can result in such a large difference in the inferred galaxy properties (see more discussion in \citealt{Naidu2022ApJ...940L..14N,Naidu2022arXiv220802794N}).

The halo mass evolution suggests that a number of merger events might have occurred in the formation history of galaxies like GLz12. For this particular analogue, its progenitor merges into a more massive halo at $z{\sim}13$, which introduces a significant increase in its star-forming gas component and triggers a re-ignition of star formation at a rate of more than $1{\rm M}_\odot {\rm yr}^{-1}$. Late on, around $z{=}12.2$, the analogue encounters another major merger, further boosting star formation activities and reaching an SFR of ${\gtrsim}10{\rm M}_\odot {\rm yr}^{-1}$. These also lead to significant fluctuations in the predicted galaxy size (around $z{\sim}12.2$) between the measured values of ${\sim}0.5$ pkpc \citep{Naidu2022ApJ...940L..14N} and $0.06{\pm}0.01$pkpc \citep[][]{Ono2022arXiv220813582O}, which chose different point spread functions and images during the analysis. 

Note that in our model, star-forming discs are assumed to be rotationally supported, conserving specific angular momenta when the gas cools from an initially virialized state, and following an exponential surface density profile. These assumptions, made by many theoretical models (e.g. \citealt{Henriques2015MNRAS.451.2663H,Stevens2016MNRAS.461..859S}) facilitating evaluation of galaxy sizes from their host halo properties, have been shown successful when predicting low-redshift observations -- the scale radius is $R_{\rm s}{=}R_{\rm vir}(\lambda/\sqrt{2})$ where $R_{\rm vir}$ and $\lambda$ are the virial radius and halo spin parameter while the effective radius (or half-light radius) is $R_{\rm e}{\sim}1.68R_{\rm s}$. However, the increasing merger rate towards higher redshifts implies that galaxies may not have enough time to recover from previous merger events despite having shorter dynamical timescales \citep{Poole2016MNRAS.459.3025P}. Although GLz12 shows no sign of multiple clumps down to 0.05 kpc, there are an increasing number of observations suggesting that galaxies at high redshift do not have a simple disc-like morphology and present signs of interaction (e.g. \citealt{Treu2022arXiv220713527T,Witten2023arXiv230316225W,Whitler2023arXiv230516670W}). Therefore, we caution against over-interpreting our prediction of galaxy sizes.

The fate of the GLz12 analogue is to steadily increase its UV luminosity and the stellar content with a stellar-to-halo-mass ratio that increases from 0.5 per cent at $z{=}12.2$ to 5 per cent at $z{\sim}6$. It is evident from Fig. \ref{fig:glf_z12_analogues_histories} that despite the bursty nature of star formation in GLz12's analogue, its extremely bright UV radiation is not transient. For comparison, we show the property histories for the next 5 most luminous galaxies identified at $z{=}12.2$. More than half of these galaxies become much fainter than the GLz12 analogue at later times with some even dropping luminosities for ${\gtrsim}100$Myr after $z{=}12$.

Finally, in agreement with the expectation (e.g. \citealt{Bouwens2010ApJ...708L..69B,Bouwens2014ApJ...793..115B}) for galaxies with a UV continuum slope of ${\sim}$-$2.3\pm0.1$ \citep[][see also \citealt{Cullen2022arXiv220804914C} for a larger JWST sample]{Naidu2022ApJ...940L..14N}, our model also suggests that the GLz12 analogue experiences negligible dust attenuation at $z{\gtrsim}12.2$. This is mainly driven by its low metallicity\footnote{Mass and radius of galaxies also play a role in determining dust attenuation according to our model \citep{Qiu19}. However, the impact from differences between the predicted and measured galaxy sizes can be minimized by a normalization factor, which was calibrated against the observed UV luminosity function and colour across a large magnitude range.}, which is only ${\sim}10$ per cent of the solar value and aligns with recent ALMA follow-up finding no strong [OIII] emission from GLz12 \citep[][see also \citealt{Popping2023A&A...669L...8P}]{Bakx2023MNRAS.519.5076B}. The theoretical interpretation for such massive galaxies experiencing little dust attenuation is that either they have ejected their dust contents or current star-forming clouds are segregated from dust that was generated in earlier episodes of star formation (see e.g. \citealt{Ziparo2022arXiv220906840Z}). As the gas fraction of the GLz12 analogue remains high at $z{\sim}12$, it is likely that UV emitting regions and dust are indeed of different origins in high-redshift galaxies (see e.g. \citealt{Behrens2018MNRAS.477..552B,Sommovigo2020MNRAS.497..956S}). In fact, the star-forming disc only reaches the level of solar metallicity at $z{\sim}8$ where the optical depth to UV non-ionizing photons inside the birth cloud of stars exceeds $\tau_{1600}=1$. At this point, the cloud can absorb a significant fraction of UV photons (e.g. compare the thick and thin coloured lines in panel 1 of Fig. \ref{fig:glf_z12_analogues_histories}) before having dissipated after 10Myr \citep{Charlot2000ApJ...539..718C}. UV photons also experience attenuation by the diffuse ISM dust although our model suggests this only becomes significant at very late times ($z{\sim}6$).

\subsection{Maisie’s Galaxy} \label{CEERSz14}

\begin{figure*}
	\centering
	\includegraphics[width=.87\textwidth]{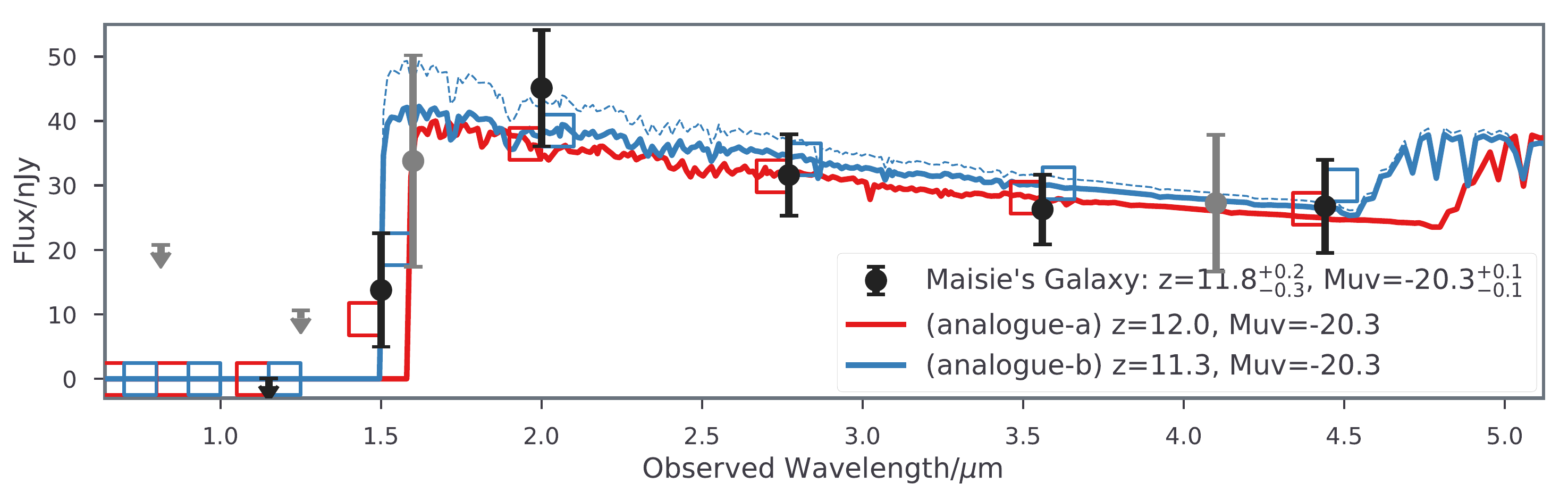}\\\vspace*{-2mm}
	
	\begin{minipage}{0.37\linewidth}
		\includegraphics[width=\linewidth]{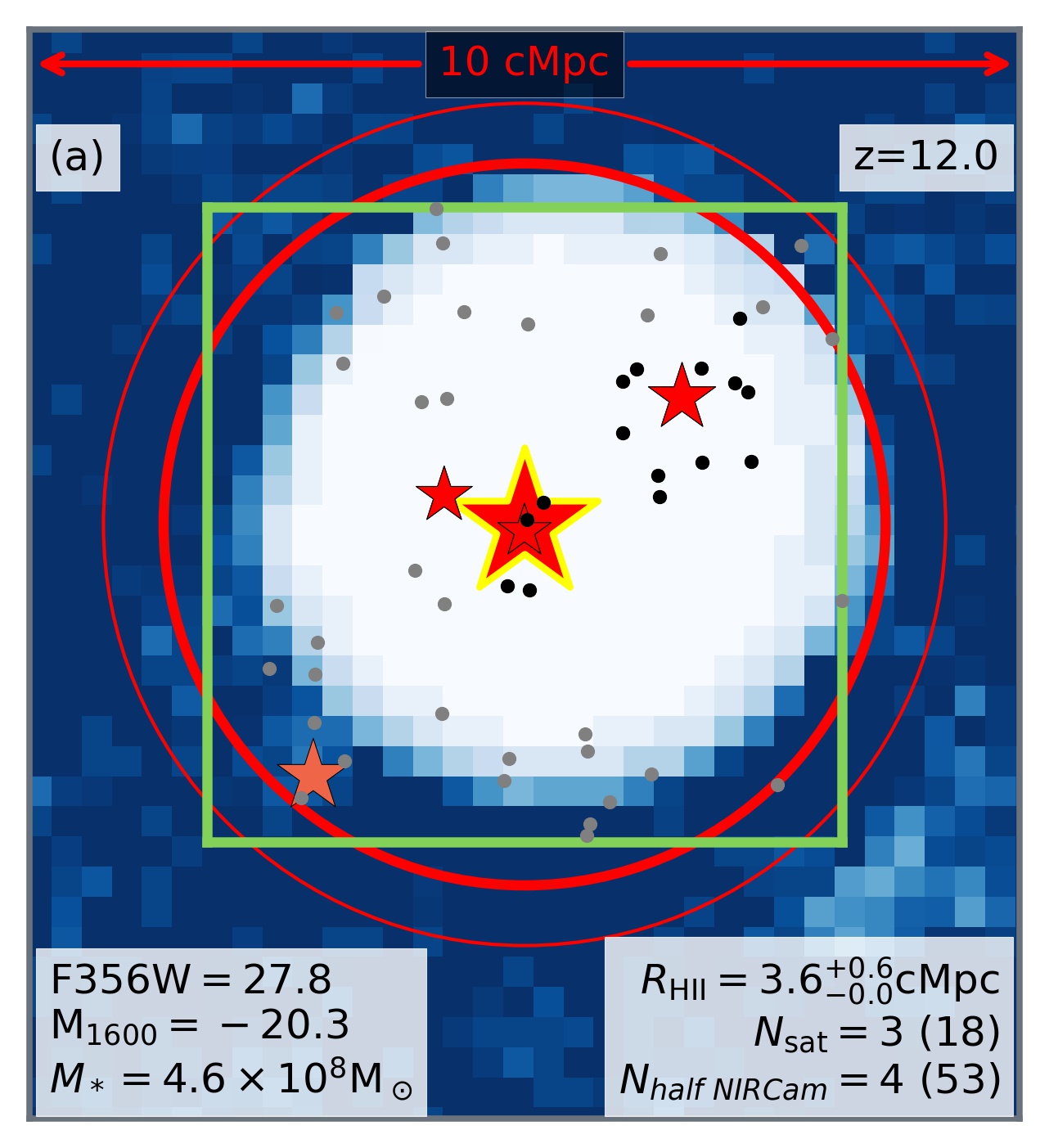}\\
		\includegraphics[width=\linewidth]{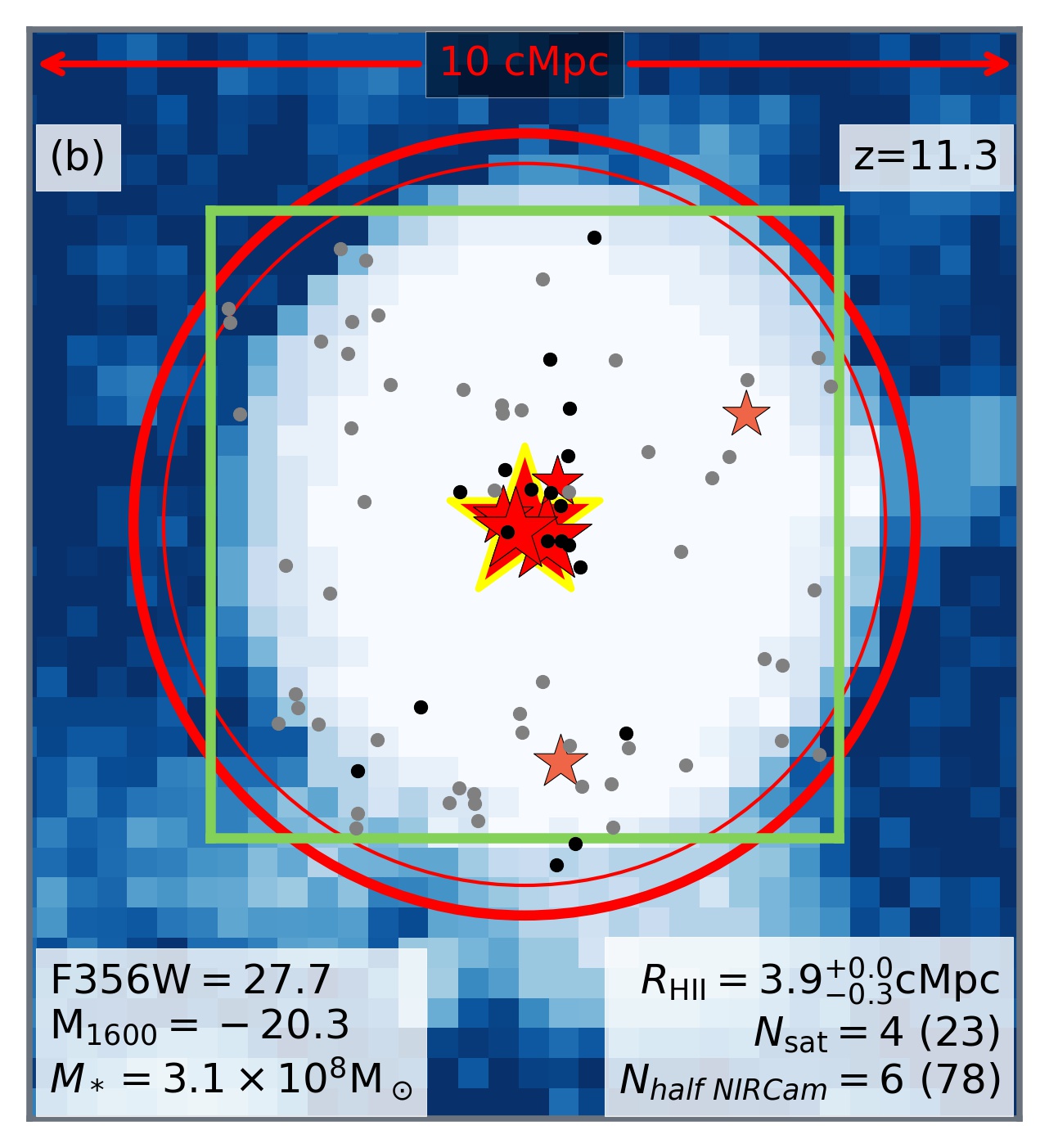}		
	\end{minipage}\hspace*{-2mm}
	\begin{minipage}{0.2\linewidth}
		\begin{minipage}{\linewidth}
			\includegraphics[width=\linewidth]{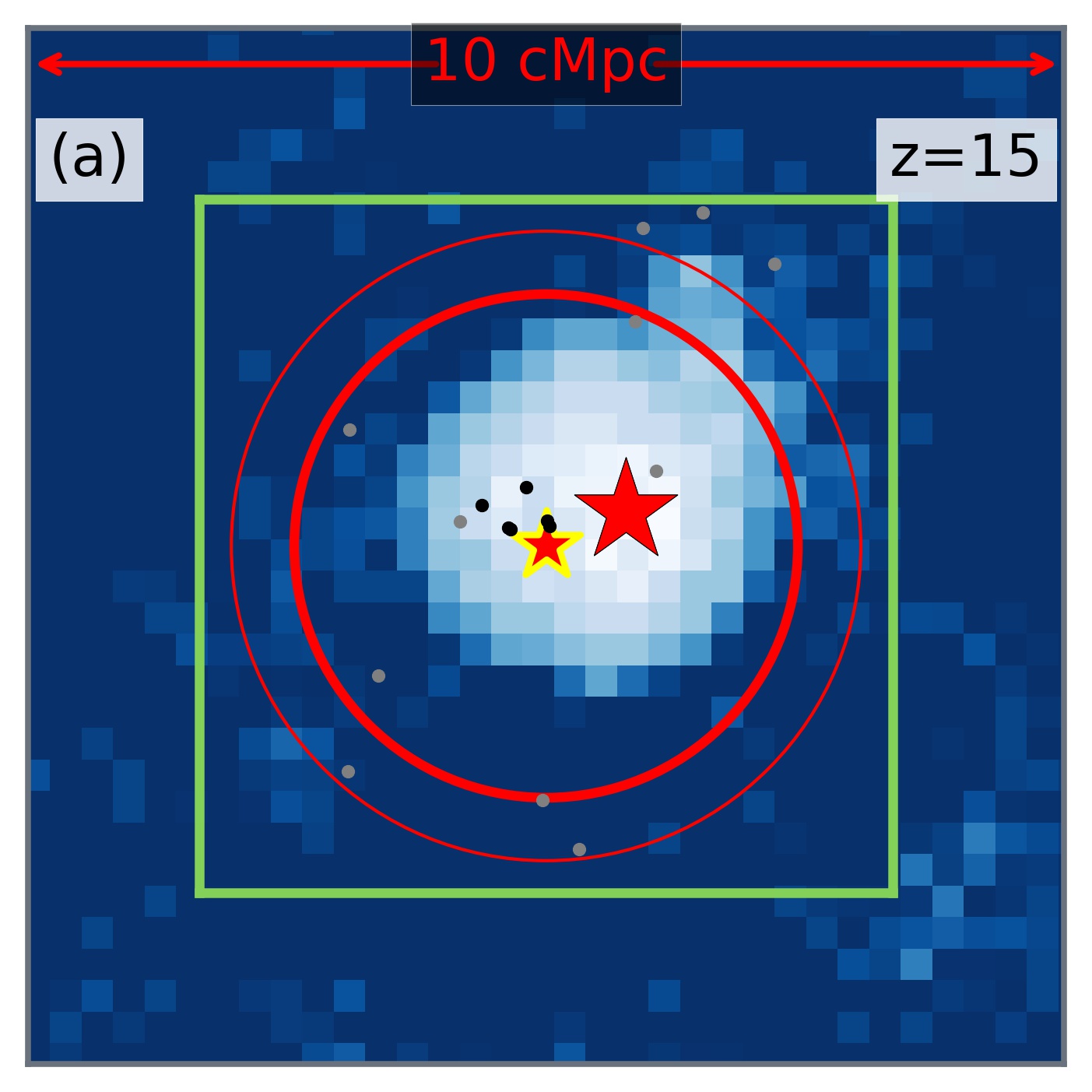}\\
			\includegraphics[width=\linewidth]{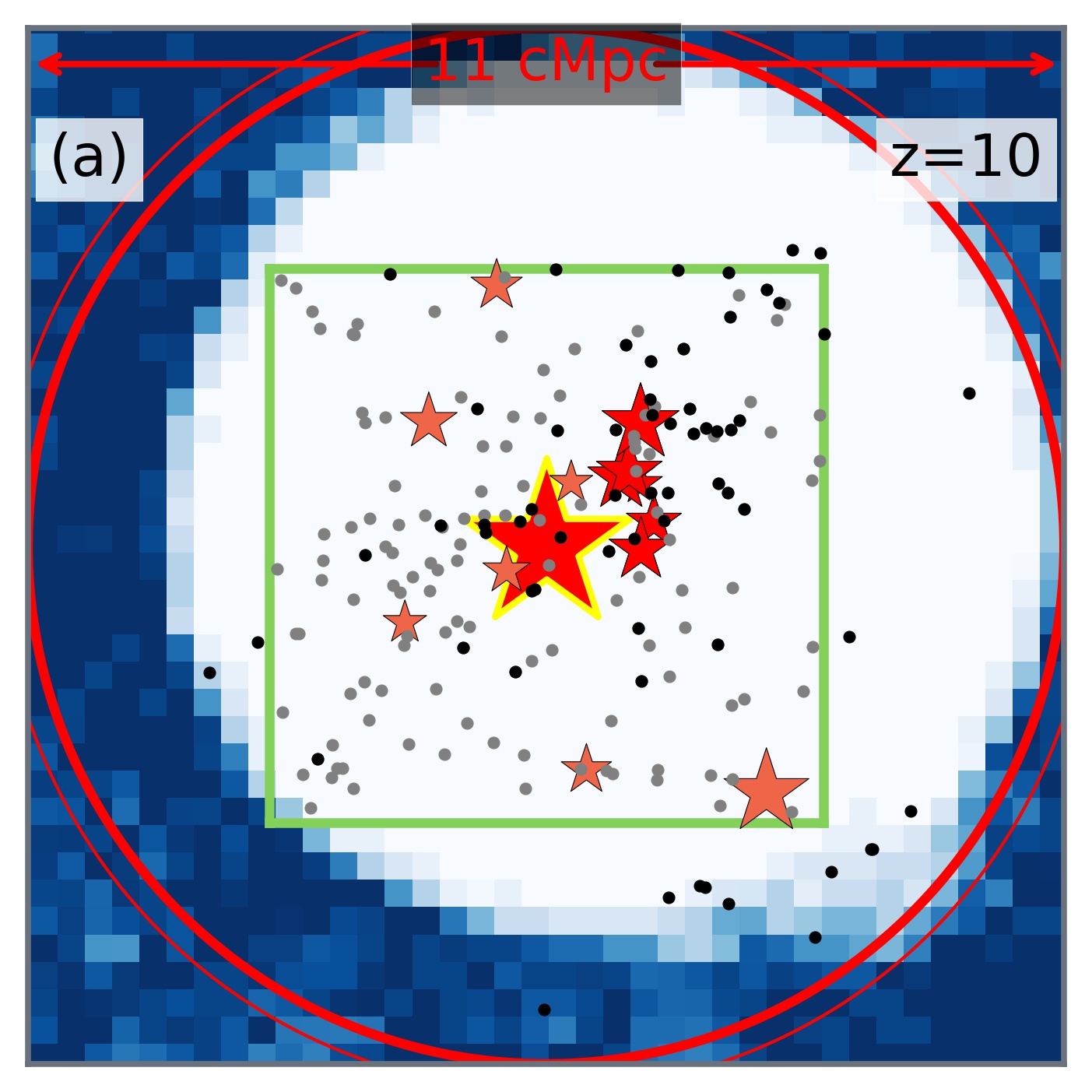}\\
			\includegraphics[width=\linewidth]{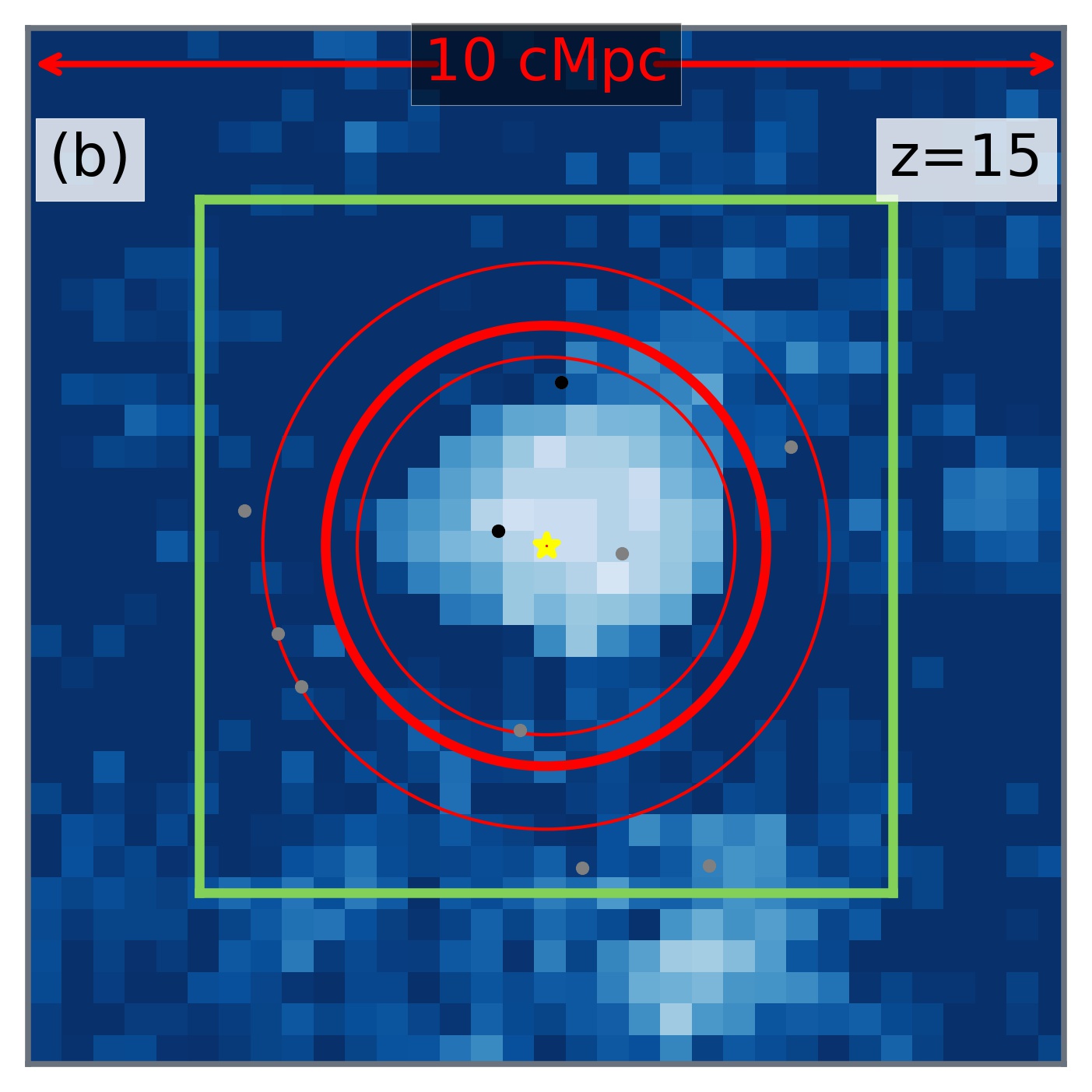}\\				\includegraphics[width=\linewidth]{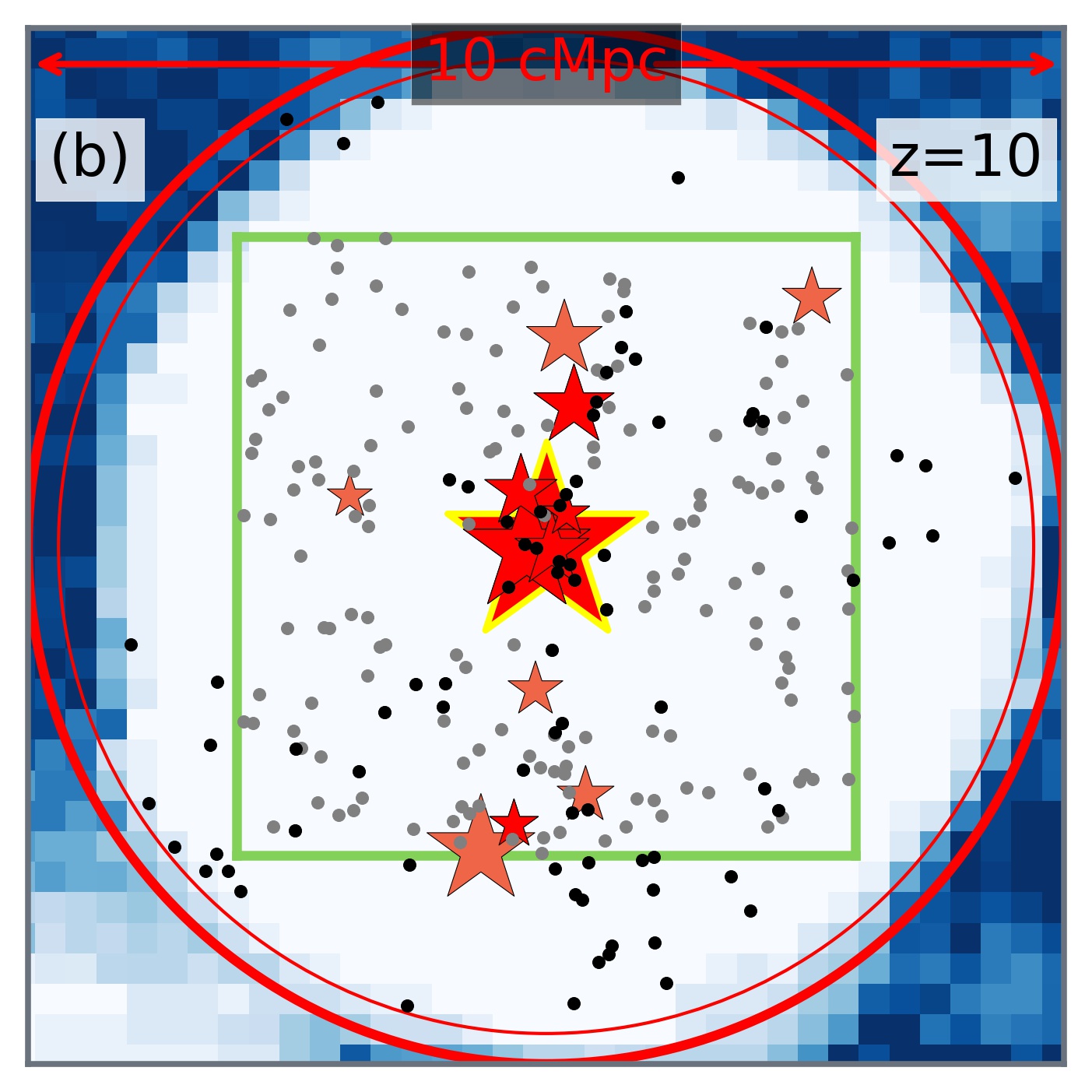}
		\end{minipage}
	\end{minipage}
	\begin{minipage}{0.405\linewidth}	
		\includegraphics[width=\linewidth]{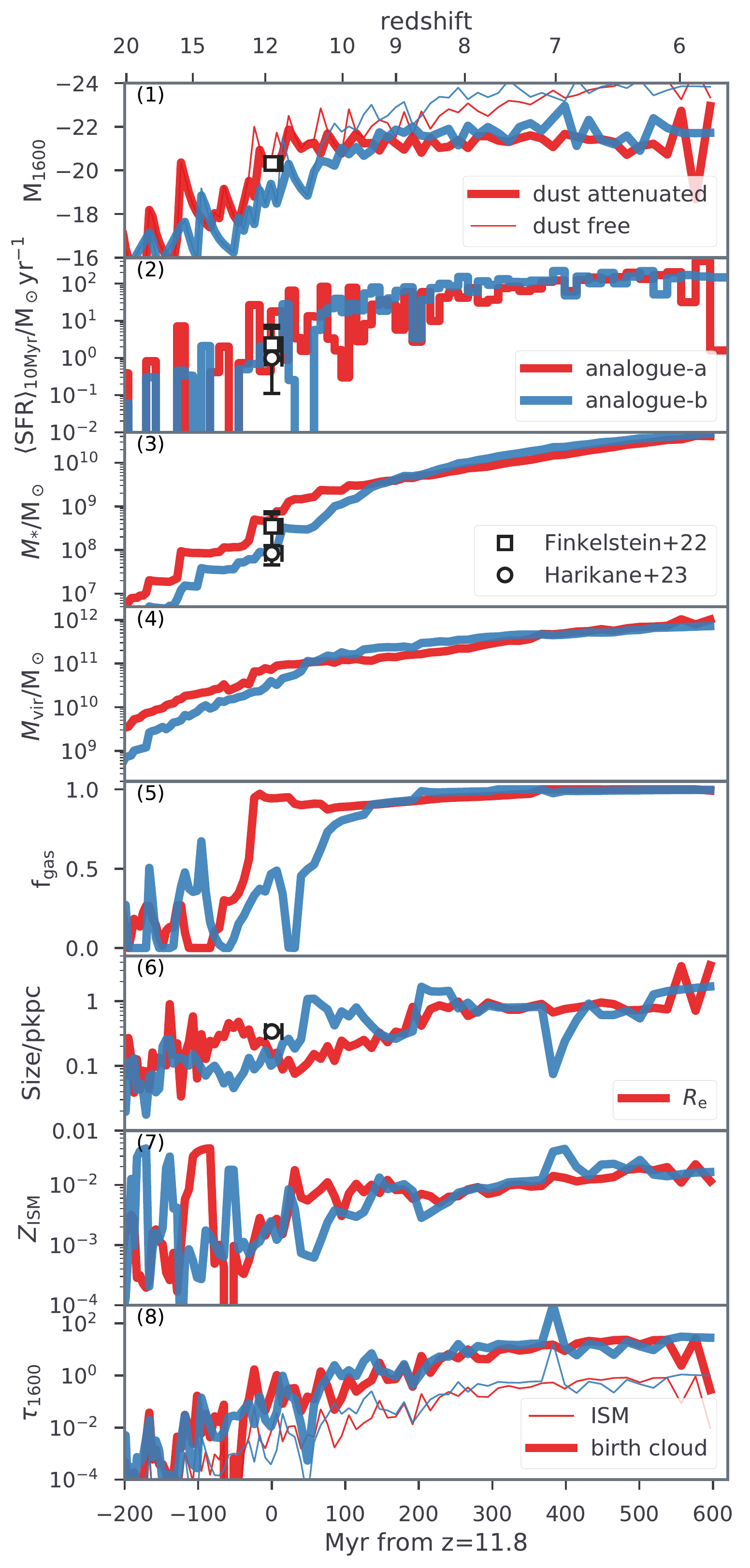}
	\end{minipage}\vspace*{-1mm}
	\caption{Maisie's Galaxy analogues. See captions of Figs. \ref{fig:glf_z12_analogues_sed} \& \ref{fig:glf_z12_analogues_histories}  for more plot details. However, two example analogues are presented here with additional photometry from HST and JWST F410M (still consistent with the analogue despite not being included during selection) indicated in grey in the top pane. Also note that SFR at $z{\gtrsim}11.5$ is averaged over several snapshots with total intervals of ${\sim}10$ Myr following \citet{Finkelstein2022ApJ...940L..55F} while at later times it is averaged over one snapshot with an increasing time step from 10 to 20Myr. Observational results are taken from \citet{Finkelstein2022ApJ...940L..55F} and \citet{Harikane2023ApJS..265....5H}.
	}
	\label{fig:ceers_z14_analogues_sed}
\end{figure*}

Maisie's Galaxy at $z=11.44^{+0.09}_{-0.08}$ was reported by multiple teams including \citet{Finkelstein2022ApJ...940L..55F}, \citet{Donnan2023MNRAS.518.6011D}, \citet{Harikane2023ApJS..265....5H} and \citet{haro2023spectroscopic} showing consistent inferred properties\footnote{Maisie's Galaxy was initially considered to be at  $z{=}14.3^{+0.4}_{-1.1}$\citep{Finkelstein2022ApJ...940L..55F}, and therefore became the second object that we looked into as it had an even higher redshift than GLz12 which was previously thought at $z{\sim}13$. Using the SED from the first version of its pre-print, we identified 2 analogues showing similar evolution and environment as the analogue presented here but with a burstier SFH.}. Using its photometric data (e.g. \citealt{Finkelstein2022ApJ...940L..55F}; $z=11.8^{+0.2}_{-0.3}$), it is estimated that Maisie's Galaxy possesses a low SFR of ${\sim}2{\rm M}_\odot {\rm yr}^{-1}$, a stellar mass around $10^{8.5}{\rm M}_\odot$, an intrinsic UV magnitude of $-20.3$, a steep UV slope of $-2.5$, a low dust extinction of $A_{v}{\sim}0.1$, and an effective radius around 0.34 pkpc. Given its lower luminosity, identifying analogues for this galaxy should be less challenging than GLz12.

\subsubsection{Analogue selection}
Within our 31 simulation snapshots between $z{=}10.8$ and 16.9, 7 galaxies are identified having fluxes consistent within at least 2$\sigma$ of the observed photometry \citep{Finkelstein2022ApJ...940L..55F} in the filters F150W, F200W F277W, F356W, and F444W, as well as being lower than the 2$\sigma$ upper limit in the non-detection band  F115W. Although we do not forward model luminosities from the HST or JWST F410M filters, the spectra of our analogues are found also consistent with these measurements/upper limits.

Our analogues share similar physical properties with the observations, including an intrinsic UV magnitude around $-20.3$, a stellar mass of ${\sim}10^{8.5}{\rm M}_\odot$, and a redshift between 11 and 12. To be concise, only two example analogues, dubbed analogue-a and b, are presented here which are found at $z{=}11.3$ and 12.0. From the top panel of Fig. \ref{fig:ceers_z14_analogues_sed}, we see the SED fitting is well-performed overall with a slightly flatter predicted spectrum compared to the observation. We notice the same challenge in fitting the UV slope was also faced by \citet[][fig. 4]{Finkelstein2022ApJ...940L..55F} and \citet[][fig. 8]{Harikane2023ApJS..265....5H} while \citet[][fig. A6]{Donnan2023MNRAS.518.6011D} found a better fit at $z=12.3$ instead. However, there are still differences between these observational results -- for instance,  F200W is measured to be 27.3 mag by \citet{Finkelstein2022ApJ...940L..55F} and 27.8 mag by the other two groups while the colour F200W-F356W is -0.4 in \citet{Donnan2023MNRAS.518.6011D} unlike the -0.6 by the rest of the teams. These might explain why Maisie's Galaxy is reported to be brighter by \citet{Finkelstein2022ApJ...940L..55F} and less blue by \citet{Donnan2023MNRAS.518.6011D}. Nevertheless, such a steep UV slope is consistent with most high-redshift star-forming galaxies having low metallicities \citep{Bouwens2014ApJ...793..115B}, aligning with the properties of our analogues which suffer little dust attenuation. 

We note that \cite{Zavala2023ApJ...943L...9Z} reported no detection of Maisie's Galaxy in a number of far-infrared and millimetre observations such as SCUBA-2, Spitzer and Hershel, and hence ruled out the scenario of strong dust emission. Moreover, while preparing this manuscript, \cite{haro2023spectroscopic} presented NIRSpec result of Maisie's Galaxy which verifies its cosmic origin -- both the spectroscopic redshift ($z=11.44_{-0.08}^{+0.09}$) and inferred galaxy properties remain consistent with the photometric results.

\subsubsection{Local environment and evolutionary paths}\label{maisie_env}

Interestingly, despite being fainter than the GLz12 analogue presented in Section \ref{gnlz12}, both analogues of Maisie's Galaxy are located in slightly larger ionized bubbles of a radius around 3.6 cMpc. This implies a dense local environment for these two analogues, which is evident in the central and lower left panels of Fig. \ref{fig:ceers_z14_analogues_sed}. We see that analogues-a and b have crowded local environment at $z{\sim}11.8$ with 3 or 4 galaxies brighter than 30.7 mag in F35W within the corresponding \textsc{Hii} bubbles (c.f. zero in the lower left panel of Fig. \ref{fig:glf_z12_analogues_sed}). A deep photometric follow-up would identify ${\sim}5$ or 50 depending on whether the field is lensed.

Looking further at their evolutionary histories, we see analogue-a is in fact a satellite galaxy at $z{=}15$ with the central galaxy having already ionized the surrounding IGM. The two are likely undergoing merger at $z{\sim}13$ with our analogue having its mass stripped first before the final merger. This is indicated by a 50 Myr trough in the halo mass history. The merger triggers an influx of star-forming gas, leading to a subsequent star formation rate (averaged over 10 Myr following \citealt{Finkelstein2022ApJ...940L..55F}) of ${\sim}20{\rm M}_\odot {\rm yr}^{-1}$ and leading to a much higher luminosity than observed. By $z{=}12$ (see the lower right panels of Fig. \ref{fig:ceers_z14_analogues_sed}), at which we identify analogue-a, its SFR drops to ${<}1{\rm M}_\odot {\rm yr}^{-1}$ and luminosity becomes more consistent with Maisie's Galaxy. On the other hand, analogue-b has a much younger stellar age as most of its stars are formed in a burst at $z{=}11.3$ with an SFR of nearly $30{\rm M}_\odot {\rm yr}^{-1}$. This burst costs all star-forming gas that analogue-b has gradually accumulated over the past 100 Myr, during which its SFR is kept low at ${<}1{\rm M}_\odot {\rm yr}^{-1}$. Moreover, the predicted galaxy size (${\sim}0.25$ pkpc) and metallicity (on the order of 0.001 to 0.01) are similar and consistent with what is suggested by the observation. 

As for the potential subsequent evolution of Maisie's Galaxy, properties of the two analogues diverge after $z{\sim}12$. As analogue-b has consumed all of its star-forming gas, its subsequent star formation becomes quenched. On the other hand, analogue-a remains highly efficient in forming stars and its stellar component at $z{\sim}11$ reaches nearly an order of magnitude larger mass than that of analogue-b. At $z{\sim}11$, a major merger event happens to analogue-b, bringing it to a similar evolutionary path as analogue-a from then on and both of them keep forming stars at a high level of  10-100${\rm M}_\odot {\rm yr}^{-1}$ until $z{\sim}6$. 

\subsection{SMACS\_z16a \& SMACS\_z16b}\label{subsection:smacz16ab}

\begin{figure*}
	\centering
	\includegraphics[width=.9\textwidth]{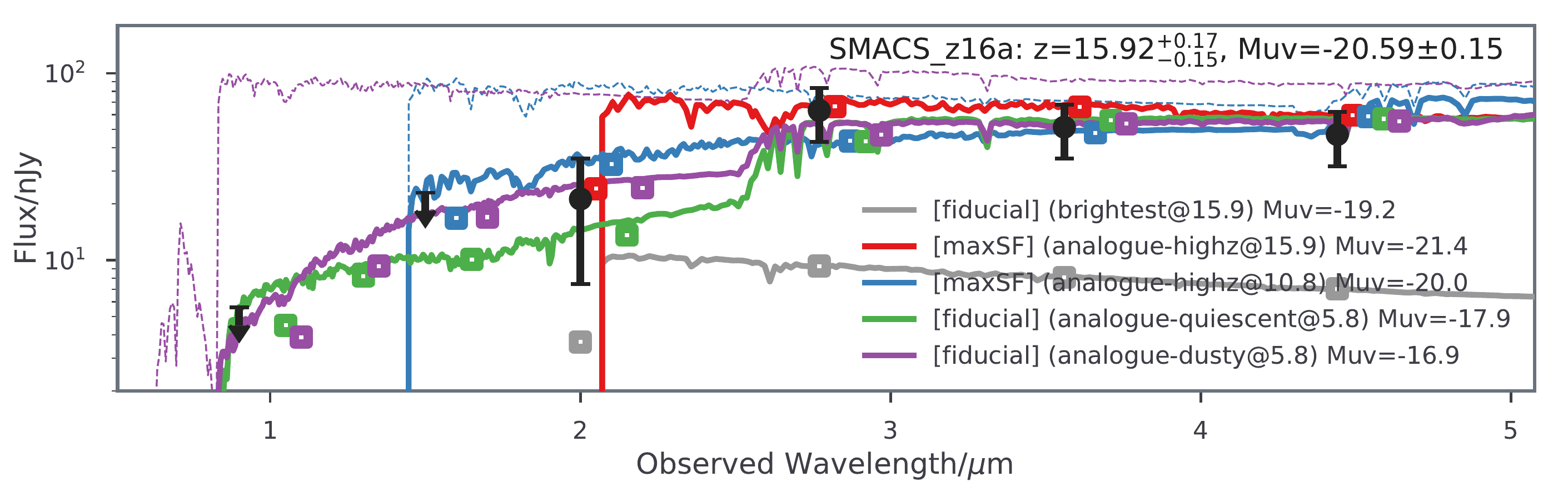}\\
	\includegraphics[width=\textwidth]{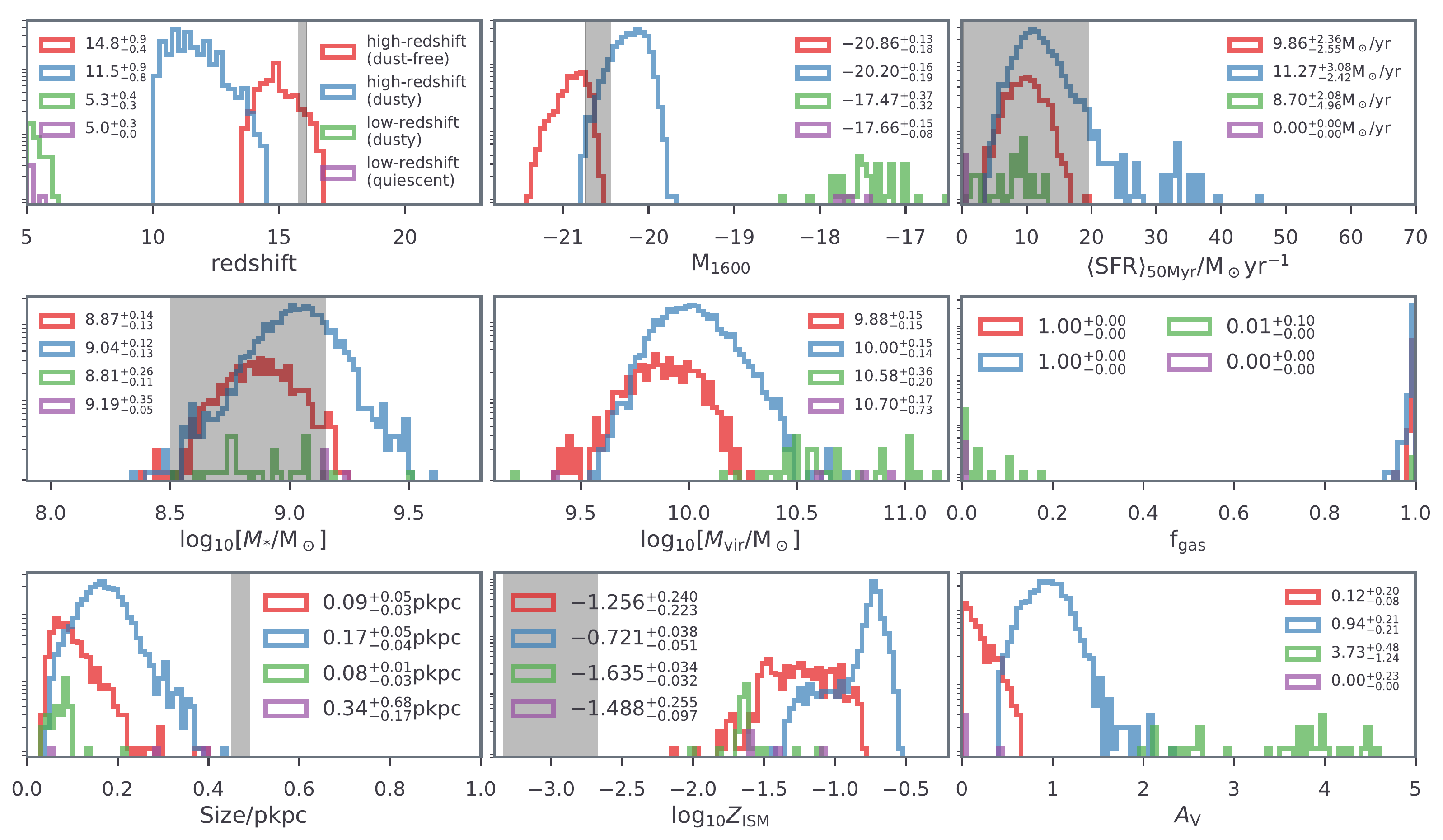}\\
	\caption{Analogues of the $z{\sim}16$ candidate from \citet[][ID: SMACS\_z16a]{Atek2023MNRAS.519.1201A}. \textit{Top panel:} we show examples for four different scenarios including quiescent and dusty galaxies at $z{\sim}6$ in the fiducial as well as high-redshift, star-forming counterparts with or without noticeable dust attenuation at $z{\sim}10$--17 in \textit{maxSF} where feedback is turned off and star formation efficiency is maximized. We additionally show the brightest galaxy in the fiducial model for comparison. See the caption of Figs. \ref{fig:glf_z12_analogues_sed} and \ref{fig:ceers_z14_analogues_sed} for more figure details, but here we offset the central wavelength of the modelled SEDs for better visualization. \textit{Bottom panels:} distribution of galaxy properties including redshift; intrinsic UV magnitude; star formation rate averaged over 50Myr; stellar mass; halo mass; the fraction of gas available for forming stars; size; metallicity; and dust extinction parameters for 471 high-redshift (dusty) analogues, 2415 high-redshift (dust-free) analogues, 30 low-redshift (dusty) analogues and 4 low-redshift (quiescent) analogues. Median values and [16,84] percentiles are presented in the top corner of each subpanel with estimated intrinsic UV magnitude, stellar mass and size from \citet{Atek2023MNRAS.519.1201A} and SFR from \citet{Furtak2023MNRAS.519.3064F} indicated by shared regions.}
	\label{fig:smacs_z16a_lowz_analogues_prop} 
\end{figure*}

In the field of SMACS, \citet[][]{Atek2023MNRAS.519.1201A} identified two galaxies at $z{\sim}16$ -- SMACS\_z16a and SMACS\_z16b (see also \citealt{Adams2023MNRAS.518.4755A,Harikane2023ApJS..265....5H}), which have intrinsic UV magnitudes of around $-20.5$. To account for gravitational lensing, we also demagnify their observed SEDs by a factor of 2.18 and 1.13 and update the uncertainties to further incorporate errors of the lensing models. The exact values are taken as the average magnification among different strong lensing models based on \citet{Furtak2023MNRAS.519.3064F}.

\subsubsection{Analogue selection}

As we have argued in Section \ref{subsection:glfz12} using the estimated number density of $z{\sim}16$ candidates, these candidates are too bright to remain consistent with our fiducial model. Therefore, we instead seek star-forming analogues in the \textit{maxSF} output in this section. However, as these candidates are still subject to spectroscopic confirmation and may be low-redshift quiescent or dusty galaxies (see e.g. \citealt{Naidu2022arXiv220802794N,Harikane2023ApJS..265....5H,haro2023spectroscopic}), we also present low-redshift analogues \textit{in the fiducial output} for comparison. 

We apply the same criteria when performing high- or low-redshift analogue searches -- the modelled SED has to be within the $2\sigma$ observational uncertainties for filters above the break (F200W, F277W, F356W, F444W) and lower than the $2\sigma$ flux threshold for non-detection (F090W \& F150W). However, the redshift range for the star-forming analogue search is extended to 57 snapshots between $z{=}10$ and 23 while for the low-redshift search, it is limited to $z{\ge}5$ as our {\it N}-body simulation has not reached later times yet.

From \textit{maxSF}, we identify 2886(505) high-redshift galaxies possessing similar SEDs as the observed one for SMACS\_z16a(b). On the other hand, when looking at the fiducial model, only 34(3) low-redshift analogues are found at $z{\ge}5$. As these two $z{\sim}16$ candidates possess qualitatively similar SEDs, we only discuss analogues of SMACS\_z16a, and present SED examples of its analogues as well as their distribution as functions of various properties in Fig. \ref{fig:smacs_z16a_lowz_analogues_prop}.

\subsubsection{High-redshift solutions}
We identify two modes for the high-redshift star-forming analogues found in \textit{maxSF} (see also Figs. \ref{fig:JADEz13} and \ref{fig:JADEz12}). Therefore, we further split the sample (approximately) based on redshift and the dust extinction parameter ($A_V$)\footnote{Moderate degeneracy between redshift, dust extinction, UV magnitude and metallicity is present in the high-dimensional distribution.}. This results in a $z{\sim}11.5$ population with $A_V{\sim}1$, which we consider as high-redshift, dusty galaxies; while the second group is centred at $z{=}15$ with very little dust attenuation. For SMACS\_z16a, these two solutions correspond to the different redshifts inferred by \citet[][$z=15.92^{+0.17}_{-0.15}$]{Atek2023MNRAS.519.1201A} and by \citet[][$z=10.61^{+0.51}_{-8.55}$]{Harikane2023ApJS..265....5H}, with the latter rejecting SMACS\_z16a as being $z{\sim}16$ because the colour F200W-F277W is not red enough. This is further illustrated by the example high-redshift, dusty analogue shown in the top panel of Fig. \ref{fig:smacs_z16a_lowz_analogues_prop}, whose F200W flux sits on the $2\sigma$ upper limits of the observation while F277W is closer to the lower threshold. 

As expected, these high-redshift analogues are forming stars at very high rates (${\sim}10{\rm M}_\odot{\rm  yr}^{-1}$), and have managed to build a relatively large stellar content with a stellar-to-halo mass ratio of nearly 10 per cent (i.e., $M_*{\sim} 10^9{\rm M}_\odot$ and $M_{\rm vir}{\sim}10^{10}{\rm M}_\odot$). As they represent areas where the first episode of star formation occurs in our universe {\it which assumes negligible stellar feedback}, they have been able to convert all accreted gas into stars ($f_{\rm gas}{\sim}1$), fuelling long-lasting star-forming events. However, this also leads to an over-prediction of the ISM metallicity compared to results from \citet{Furtak2023MNRAS.519.3064F}. This is because in {\it maxSF}, while supernovae do not provide thermal or kinetic feedback to remove metals (and gas) from the host galaxy, they continue polluting their environment. As our dust attenuation model assumes the optical depth increases towards later times and scales (almost) linearly with the metallicity (see more in \citealt{Qiu19}), this results in the two SED solutions we see here -- while the higher-redshift analogues prefer lower metallicities and are dust-free, the lower-redshift most likely exhibit an opposite trend. 

\subsubsection{Low-redshift solutions}
The low-redshift analogues found in our fiducial model can also be divided into two groups -- based on the extinction parameter,  galaxies with $A_v>0.2$ are considered to be dusty star-forming galaxies and isolated from quiescent counterparts. Example SEDs and property distributions for these two scenarios are shown in Fig. \ref{fig:smacs_z16a_lowz_analogues_prop} for comparison. 

Although it is a small sample\footnote{The small sample is a result of the limited redshift prior imposed by our parent {\it N}-body simulation, in particular for the quiescent galaxy scenario. In addition, using the earlier pre-print version of \citet{Atek2023MNRAS.519.1201A} that has higher luminosities than the publication, we found hundreds of low-redshift analogues with much distinct and statistical differences between dusty and quiescent galaxies and from their high-redshift, star-forming counterparts.}, we see that the low-redshift analogues are on average 3 magnitudes fainter than the high-redshift cases, with a much lower stellar-to-halo mass ratio of around 2 per cent -- they are galaxies with similar stellar mass (${\sim}10^9 {\rm M}_\odot$) but inside much larger halos (${\sim}5{\times}10^{10}{\rm M}_\odot$). Compared to the dusty high-redshift scenario, the dusty analogues at low redshift are also forming stars at ${\sim}10 {\rm M}_\odot {\rm yr}^{-1}$ but out of a relatively smaller gas content (mostly $f_{\rm gas}{\lesssim}$20 per cent) as a result of supernova feedback implemented in standard galaxy-formation modelling (i.e., our fiducial). In addition, with similar disc sizes and metallicities, the low-redshift dusty galaxies suffer significant attenuation. On the other hand, the quiescent analogues have zero star formation rate, larger discs and low attenuation. Observationally, even though quiescent galaxies with such low masses are not common at $z{\gtrsim}5$, there have been tentative reports with JWST of quiescent galaxies at early times \citep{Looser2023arXiv230214155L} and lower masses \citep{Strait2023arXiv230311349S}.

\subsection{S5-z16-1}\label{subsection:z17}

\begin{figure*}
	\centering
	\includegraphics[width=.9\textwidth]{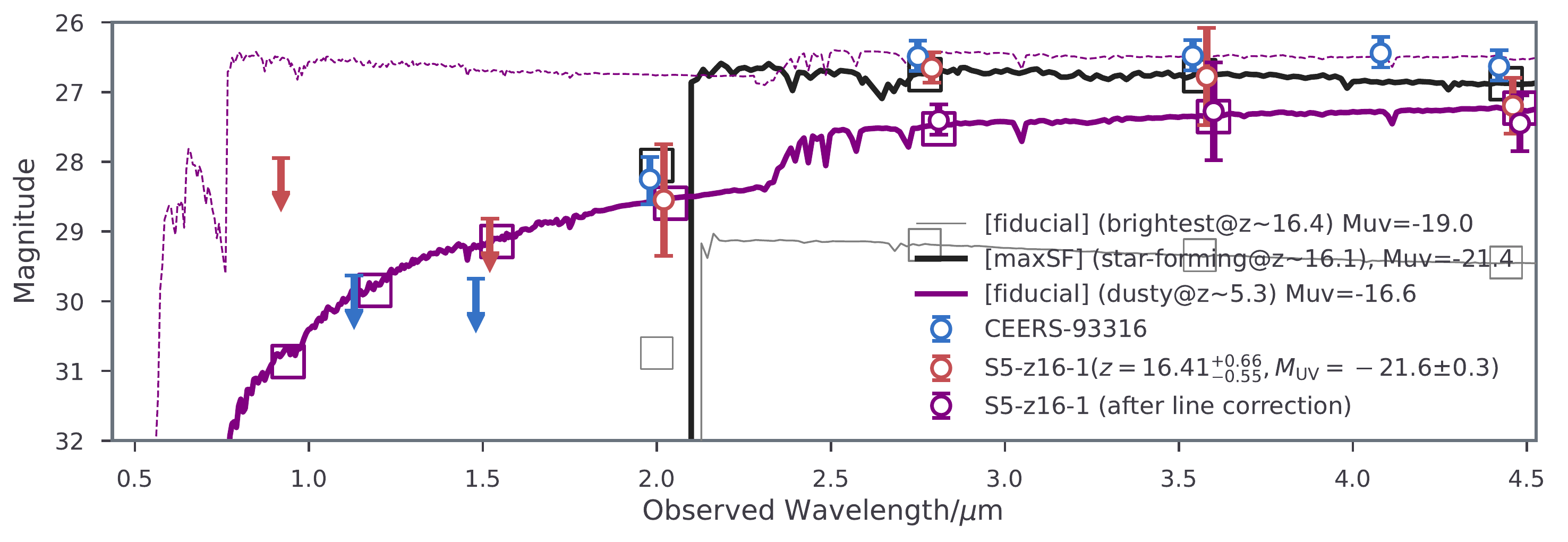}\\\vspace*{-2mm}
	\includegraphics[width=\textwidth]{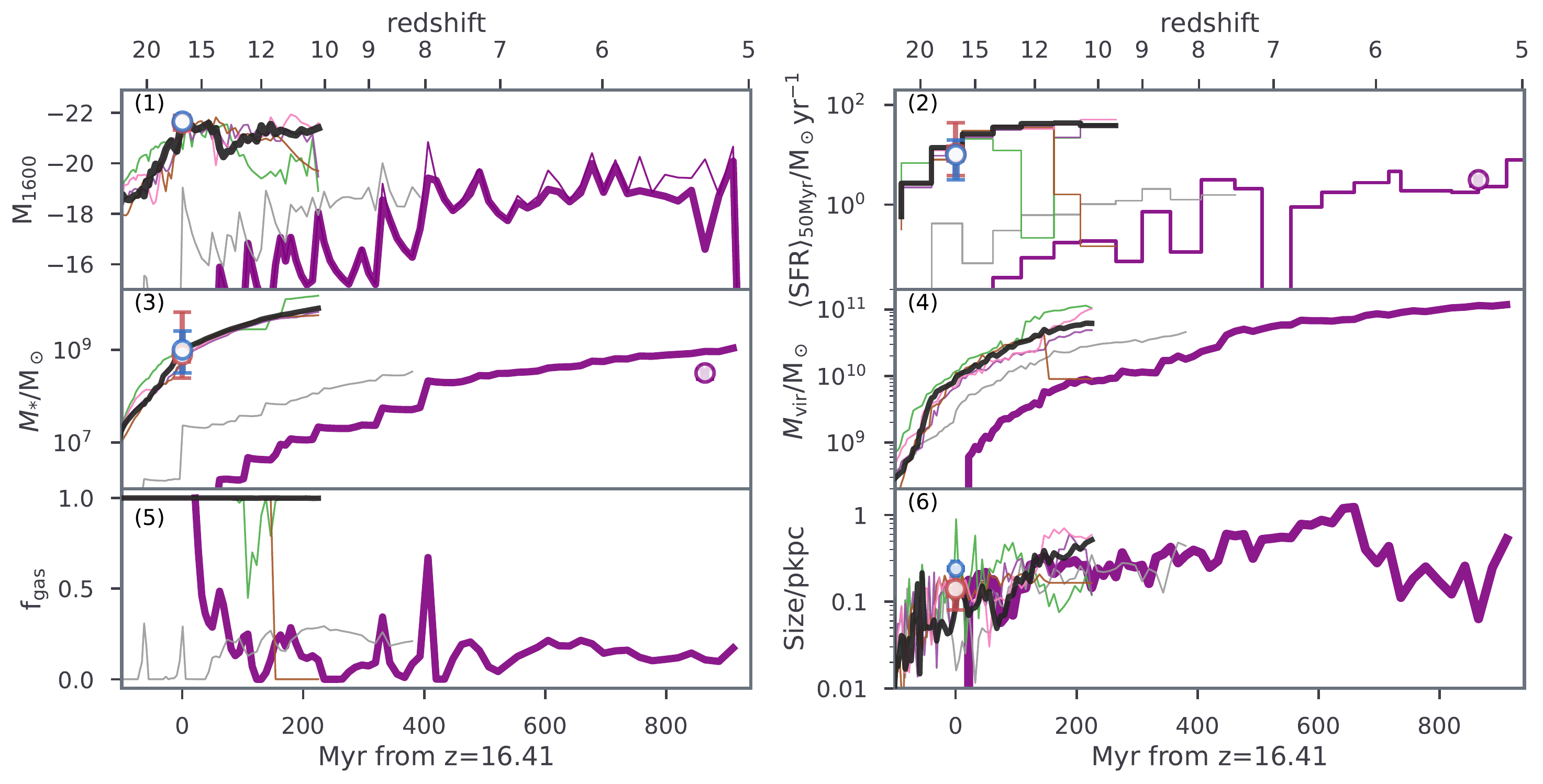}\\\vspace*{-1mm}
	\caption{\textit{Top panel:} (a) spectrum (thick black curve) and SED (black squares) of a star-forming galaxy at $z{\sim}16$ in {\it maxSF} which can be considered as an analogue for both S5-z16-1 (red circles; \citealt{Harikane2023ApJS..265....5H}) and CEERS-93316 (if it were at $z{\sim}16$; blue circles; \citealt{Donnan2023MNRAS.518.6011D,Harikane2023ApJS..265....5H,Naidu2022arXiv220802794N}). (b) spectra and SED (thick, purple) of a dusty galaxy (in the fiducial catalogue) that is analogue to S5-z16-1 {\it after line correction} (purple circles). Note that the central wavelengths of these SEDs are offset for better visualization. \textit{Bottom panels:} possible evolution as in (1) intrinsic UV magnitude; (2) star formation rate averaged over ${\sim}50$Myr; (3) stellar mass; (4) halo mass; (5) fraction of star-forming gas; and (6) galaxy disc size for analogues shown above (thick lines) with observational results from \citet{Harikane2023ApJS..265....5H} and \citet{Donnan2023MNRAS.518.6011D} also indicated for comparison (red and blue circles; also note that the two results overlap with each other and sizes are measured by \citealt{Ono2022arXiv220813582O}). Stellar mass and star formation rate of CEERS-93316 at $z{=}4.9$ obtained by \citet{haro2023spectroscopic} is shown (purple circles) as a comparison for our dusty analogue identified at $z{\sim}5.3$. We additionally show the brightest galaxy (thin grey lines) in the fiducial model as well as the remaining 4 star-forming analogues (thin coloured lines) of S5-z16-1 found in \textit{maxSF}, which stops at $z{\sim}10$.}
\label{fig:Harikane_z17_lowz_S5} 
\end{figure*}

S5-z16-1 \citep{Harikane2023ApJS..265....5H} is a $z{=}16.41^{+0.66}_{-0.55}$ candidate identified from Stephan's Quintet. With an intrinsic UV magnitude of $-21.6$, S5-z16-1 is even brighter than the two lensed candidates discussed in the previous section. Therefore, identifying its analogues becomes more challenging and the {\it maxSF} model presents only five galaxies that share its SED (selection criteria identical to Sec. \ref{subsection:smacz16ab}).

As mentioned in Sec. \ref{subsub:z16}, CEERS-93316 was also a bright $z{\sim}16$ candidate \citep{Donnan2023MNRAS.518.6011D,Harikane2023ApJS..265....5H,Naidu2022arXiv220802794N} but is now considered to be a $z{=}4.9$ dusty galaxy \citep{haro2023spectroscopic}. The NIRSpec observation reveals that strong nebular lines such as [OIII] and H$\alpha$ have boosted its NIRCam photometry (particularly for band F277W), leading to an apparent break as well as the biased interpretation of its redshift. In fact, because these two galaxies have very similar SEDs, their inferred galaxy properties (when assuming they are at $z{\sim}16$) are also very close (see data points at $z{\sim}16$ in Fig. \ref{fig:Harikane_z17_lowz_S5}). One of the five analogues we find for S5-z16-1 happens to also be the only analogue we can find for CEERS-93316 in \textit{maxSF}. Therefore, the revised interpretation of CEERS-93316 provides a warning that SED fitting, including that done in this work, requires better handling of the emission profile. 

\subsubsection{High-redshift solutions}

Figure \ref{fig:Harikane_z17_lowz_S5} shows the SED and possible evolutionary path of S5-z16-1 with a comparison against the observations. The property histories for the other 4 analogues of S5-z16-1 are also included and we see that the prediction at $z{\sim}16$ from our \textit{maxSF} model agrees very well with the observational results. The S5-z16-1 analogues are extremely bright with stellar masses around $10^9{\rm M}_\odot$ and high stellar-to-halo mass ratios of ${\sim}10$ per cent \citep{Harikane2023ApJS..265....5H}. When averaged over 50 Myr, the star formation rates are on the order of $10{\rm M}_\odot {\rm yr}^{-1}$. 

It is also remarkable that all 5 analogues show consistent formation histories before $z{\sim}13$ -- they build their stellar contents very rapidly in this 200 Myr interval starting with $M_*{\sim}10^7{\rm M}_\odot$ and $M_{\rm vir}{\sim}3{\times}10^8{\rm M}_\odot$ at $z{\sim}25$, reaching $M_*{\sim}3{\times}10^9{\rm M}_\odot$ and $M_{\rm vir}{\sim}2{\times}10^{10}{\rm M}_\odot$ at $z{\sim}13$. This is because \textit{maxSF} includes no feedback, which allows these analogues to be able to convert all their gas into stars and significantly reduces the stochasticity in the formation history. We see that from $z{\sim}13$, the evolution of these analogues starts to diverge as a result of different merger histories.

\subsubsection{Low-redshift solutions}
Using the broad-band photometry measured by \citet{Harikane2023ApJS..265....5H}, no low-redshift analogues can be identified in our fiducial output (which stops at $z{=}5$). However, based on the line correction \citet{haro2023spectroscopic} inferred for CEERS-93316, we alter the SED of S5-z16-1 by +0.75, +0.50 and +0.25 mags in F277W, F356W and F444W to account for potential contamination\footnote{We choose to not perform line correction for SMACS\_z16a(b) in Section \ref{subsection:smacz16ab} due to the highly uncertain emission strength for high-redshift dusty galaxies. However, assuming the flux of the emission line is proportional to the continuum, we expect similar levels of alteration in magnitude to SMACS\_z16a(b). Qualitatively, this improves the fitting of the two example low-redshift analogues shown in the top panel of Fig. \ref{fig:smacs_z16a_lowz_analogues_prop}.}. Using this updated SED (see purple circles in Fig. \ref{fig:Harikane_z17_lowz_S5}), we successfully identify 1225 dusty analogues between $z{=}5$ and 6 in the fiducial catalogue.

Figure \ref{fig:Harikane_z17_lowz_S5} presents an additional example analogue for low-redshift dusty galaxies (at $z{\sim}5.3$). This analogue has a dramatically different evolutionary path compared to all high-redshift analogues we have studied in this work. The example analogue is firstly identified\footnote{In this work, augmentation is not applied to halos identified at such low redshifts when reionization has already finished.} at $z{\sim}15$ with a halo mass of ${\sim}6{\times}10^8{\rm M}_\odot$ and a less than 0.1 per cent stellar content. Although bursty, it keeps forming stars at ${\lesssim}1{\rm M}_{\odot}$/yr (when averaged over 50 Myr) until $z{\sim}8$ when its stellar mass reaches $2{\times}10^{8}{\rm M}_{\odot}$ and the stellar-to-halo-mass ratio increases to 0.3 per cent. Afterwards, this galaxy quickly gains more masses and, at $z{\sim}5.3$, its halo and stellar masses become nearly $10^{11}{\rm M}_{\odot}$ and $10^{9}{\rm M}_{\odot}$, respectively. While the metallicity of this galaxy is twice the solar level at $z{\lesssim}6$, its shrinking disc size caused by a reducing halo spin makes the disc more opaque and therefore a significant amount of UV radiation becomes saturated, making this galaxy a non-detection in JWST's F090W and F150W bands (see the top panel of Fig. \ref{fig:Harikane_z17_lowz_S5}). It is worth noting that this dusty galaxy has an extinction parameter of $A_{v}{\sim}3.5$ which is consistent with typical values at $3{<}z{<}6$ (e.g. \citealt{Barrufet2022arXiv220714733B,Rodighiero2022arXiv220802825R}). 

\section{Conclusion}\label{sec:conclusion}

JWST has delivered unprecedented data of our early Universe, revealing galaxy formation in the first 300 Myr of the cosmic time. In this work, we utilize a large-volume, high-resolution cosmological simulation coupled with a semi-analytic galaxy-formation model to study the possible evolutionary paths as well as the local environment for eight JWST galaxy candidates at $z{\ge}12$. These include three faint ($M_{\rm UV}{\gtrsim}19.5$) galaxies at $z{\sim}12$ -- JADES-GS-z13, JADES-GS-z12, S5-z12-1; two bright galaxies at $z{\sim}12$ -- GLz12,  Maisie's Galaxy; and three bright galaxies at $z{\sim}16$ -- SMACS\_z16a, SMACS\_z16b, S5-z16-1  (e.g., \citealt{Atek2023MNRAS.519.1201A,Curtis-Lake2022arXiv221204568C,Finkelstein2022ApJ...940L..55F,Harikane2023ApJS..265....5H,Naidu2022ApJ...940L..14N}).

We find faint JWST galaxies to be consistent with the standard galaxy-formation model while the bright ones are challenging or inconsistent depending on their redshift. Using our fiducial model, which is statistically representative of the observed Universe across most cosmic time ($5<z<13$) and the observed magnitude range, we show
\begin{enumerate}
	\item large samples of analogues have broad-band photometry that is consistent with the faint JWST galaxies. The distribution of our modelled galaxy properties is also broadly aligned with the SED fitting results to observations.
	But due to the burstier nature of star formation predicted by our model, the inferred stellar masses are lower than observations that commonly assume a more continuous star formation history;
	\item as a result of low number density, bright $z{\sim}12$ galaxies only have a handful of analogues in the fiducial model, whose properties are similar to values obtained through inverse modelling of observed SEDs. Although a small sample, these analogues in general suggest that bright JWST targets have a rapid build-up of their stellar content and are located in dense regions with their local environment having diverse possibilities; and
	\item our fiducial simulation does not contain bright analogues for $z{\sim}16$ candidates found in the small volume of these ERS programs. However,
	the observed SED of these $z{\sim}16$ candidates can still be reproduced by low-redshift galaxies in the fiducial model, which either are experiencing strong dust attenuation of their UV radiation or have quenched star formation to exhibit a Balmer break.
\end{enumerate}

To reproduce bright $z{\sim}16$ JWST candidates, we find that highly efficient star formation with no feedback regulation is required.
The formation history of these extremely bright analogues in this model demonstrates that they have an incredibly high stellar-to-halo mass ratio that is close to the cosmic mean baryon fraction. This suggests that while feedback and regulated star formation are essential to galaxy formation during most of the cosmic time, the confirmation of $z{\sim}16$ galaxies would indicate that this was not the case for the first massive galaxies formed during the cosmic dawn.

\section*{Acknowledgements}
We thank S. Finkelstein, S. Tacchella and D. Breitman for their comments.
This research was supported by the Australian Research Council Centre of Excellence for All Sky 
Astrophysics in 3 Dimensions (ASTRO 3D), through project \#CE170100013. Part of this work was performed on the OzSTAR and Gadi national computational facilities. YQ acknowledges HPC resources from ASTAC Large Programs and the RCS NCI Access scheme as well as its Cloud at The University of Melbourne.

\section*{Data Availability}

The data underlying this article will be shared on reasonable request to the corresponding author.



\bibliographystyle{mnras}
\bibliography{example} 








\label{lastpage}
\end{document}